\documentclass[useAMS,usenatbib]{mnras}
\usepackage[dvipdfmx]{graphicx}
\usepackage{subfigure}
\usepackage{float}
\usepackage{amsmath}
\usepackage{color}
\usepackage{ulem}
\usepackage{multirow}
\RequirePackage[l2tabu, orthodox]{nag}

\title[The Impact of Galactic Disc Environment on Star-Forming Clouds]{The Impact of Galactic Disc Environment on Star-Forming Clouds}

\author[Nguyen, Pettitt, Tasker \& Okamoto]{Ngan K. Nguyen$^{1}$\thanks{ngannk@astro1.sci.hokudai.ac.jp}, Alex R. Pettitt$^1$ , Elizabeth J. Tasker$^2$ and Takashi Okamoto$^1$\\
$^{1}$Department of Physics, Faculty of Science, Hokkaido University, Kita 10 Nishi 8 Kita-ku, Sapporo 060-0810, Japan\\
$^{2}$Institute of Space and Astronomical Science, Japan Aerospace Exploration Agency, Yoshinodai 3-1-1, Sagamihara, \\Kanagawa 252-5210, Japan}

\begin{document}
\pagerange{\pageref{firstpage}--\pageref{lastpage}} \pubyear{}
\maketitle
\label{firstpage}

\begin{abstract}
We explore the effect of different galactic disc environments on the properties of star-forming clouds through variations in the background potential in a set of isolated galaxy simulations.  Rising, falling and flat rotation curves expected in halo dominated, disc dominated and Milky Way-like galaxies were considered, with and without an additional two-arm spiral potential. The evolution of each disc displayed notable variations that are attributed to different regimes of stability, determined by shear and gravitational collapse. The properties of a typical cloud were largely unaffected by the changes in rotation curve, but the production of small and large cloud associations was strongly dependent on this environment. This suggests that while differing rotation curves can influence where clouds are initially formed, the average bulk properties are effectively independent of the global environment. The addition of a spiral perturbation made the greatest difference to cloud properties, successfully sweeping the gas into larger, seemingly unbound, extended structures and creating large arm-interarm contrasts.
\end{abstract}

\begin{keywords}
hydrodynamics - methods: numerical - ISM: clouds - ISM: structure - galaxies: star formation - galaxies: structure.
\end{keywords}

%%%%%%%%%% Introduction %%%%%%%%%%
\section{Introduction}

Giant Molecular Clouds (GMCs) are the major reservoir for cold, molecular hydrogen in the interstellar medium (ISM) that fuels the formation of stars. The properties of these clouds have been well studied both in our Galaxy as well as in our nearby galactic neighbours. However, the impact different galactic environments have on the cloud properties remains unclear.  
 
In their seminal papers, \cite{Schmidt1959} and \cite{Kennicutt1998} reveal an empirical power-law relation between the surface density of the star formation rate ($\Sigma_{\rm SFR}$) and that of the gas ($\Sigma_{\rm gas}$). Such a relation is surprising, since star formation itself occurs on small, sub-parsec scales and should not necessarily care about the surface density of gas over regions kilo-parsecs in size. However, not only does this trend exist in most disc galaxies, there is a significant scatter about the median, with the gradient and normalisation of the relation varying both between galaxies and within regions of a single galaxy \citep{Bigiel2008, Roman2016, Morokuma2017}. Both these features suggest that star formation is not simply the product of the local gas density, but rather is strongly affected by forces dependent on environment that act on larger-scales.

% Proof of local-ish dependence
This dependence on environment can also be seen to play a role within an individual galaxy. Observations of the spiral galaxy M51 indicate a discrepancy in the properties of the GMCs within the spiral and inter-arm regions \citep{Koda2009, Meidt2012, Colombo2014}. The most massive GMCs with masses between $10^7-10^8$\,M$_\odot$ are considered Giant Molecular Associations (GMAs) and are only found in the spiral arms of M51, where the shear is reduced but the shear gradient is high \citep{Kim2008}. In their work on the barred spiral galaxy NGC\,4303, \citet{Momose2010} found that the star formation efficiency is about twice as high in the spiral arms compared to the bar. Such studies reinforce that simply the quantity of gas is not the only trigger for star formation.

In the extreme outer regions of our own Galaxy, beyond a galactocentric radius of 18\,kpc, the environment is quite different from the solar neighbourhood \citep{Kobayashi2008,Kobayashi2000,Yasui2008}. Observations in these distant locations reveal a typical size and mass of the clouds of just 5\,pc and $3 \times10^3$\,M$_\odot$; much smaller than the 20\,pc and $10^{5}$\,M$_\odot$ found in the inner Milky Way \citep{Roman2010,Izumi2014,Sun2017}. Comparing with observed cloud properties in M51, M33 and the Large Magellanic Cloud, \citet{Hughes2013} suggested that GMCs in both the outer Milky Way and low-mass galaxies are generally smaller and fainter than the molecular structures in the inner Milky Way. 

% Proof of no dependence 
Contrary to such apparent environmental sensitivity, good agreement is found between the properties of the GMCs in the inner Milky Way and M31 \citep{Rosolowsky2007a, Rosolowsky2005}. Aside from the variation of cloud size with galactocentric radius, typical cloud masses found in M33 hover around $10^5$\,M$_\odot$ and almost no cloud is more massive than $10^6$\,M$_\odot$ \citep{Rosolowsky2007b}. Recently, \cite{Freeman2017} identified a population of GMCs in M83 that is broadly similar to those found in the Milky Way and Local Group galaxies in both their size - linewidth and mass - size trends. Additional observational studies \citep[e.g.][]{Keto1986, Heyer2004, Oka2001, RosolowskyBlitz2005, Gratier2012, Wong2011} also suggest common properties between GMCs in molecular-rich environments. Even the effect of the spiral potential is debated, with \citet{Schinnerer2017} suggesting that the GMAs observed in M51 are actually blended gas spurs and can be separated into structures of a similar mass to inter-arm clouds. This matches simulations by \citet{Baba2017} who found no evolution sequence of GMC properties across a spiral arm, although notes that the clouds are still affected by the external pressure determined by galactic-scale features.

% Possible reasons for dependence 
Exactly why such a dependence on environment could exist is not clear. In simulations comparing the evolution of GMCs with and without star formation and stellar feedback, \citet{Tasker2015} confirmed that environment appears to the dominant factor in controlling the evolution of the star-forming gas. In their simulations, the fragmentation of the galaxy disc and subsequent interactions between GMCs dominated in determining cloud properties compared to internal forces from stellar physics (e.g. feedback). Simulations exploring the impact of a grand design morphology in M83-type barred spiral galaxies, \citet{Fujimoto2014a, Fujimoto2014b, Fujimoto2016} suggest changes in environment determine the frequency of interactions between clouds; a rate that increases in denser areas such as the spiral arms. The occurrence of mergers and tidal tails controls the small and large end of the GMCs profile distribution. The link between cloud interactions and star formation has previously been proposed by \cite{Tan2000}, who suggested that a star formation rate triggered by cloud collisions could result in a Kennicutt-Schmidt relation. Alternatively, different environments could inject turbulence into the cloud gas to resist collapse. The galactic bar region is typically dense in gas but low in star formation, which could be explained by strong shear along the bar \citep{Reynaud1998, Sheth2000, Sorai2012, Meidt2013}. Likely, all these processes play a role in influencing the production of star-forming gas but predicting the result remains opaque.

% ZE PLAN
In light of the works outlined above, it is therefore of interest to investigate how the global structure of a galaxy impacts the cloud properties within. Do effects such as the shear rate, differential rotation or non-axisymmetric spiral perturbations in a given environment impede the growth and properties of GMCs, or are they relatively unperturbed by large scale environmental changes?

In this paper we explore the impact of the galactic environment by modelling clouds forming in global disc galaxy simulations. While the gas budget of our galaxies remains similar, we vary the environment by adjusting the surrounding potential to create a diversity of galactic rotation curves. Disc galaxies are observed to have a range of different mass models, which give rise to a wealth of different rotation curves. A large ensemble of rotation curves is seen in both observations \citep{Rubin1980,Swaters2009,Sofue2016} and cosmological simulations \citep{Stinson2010,Oman2015,Santos2016}, though considerable effort is still being invested in reproducing such a wide diversity as seen in observations \citep{Read2016,Creasey2017}.

The rotation curve of a galaxy is principally determined by the gravitational potential contributions from the dark matter halo, stellar disc and bulge, in addition to non-axisymmetric perturbations such as grand design spiral arms and inner bars \citep{Wada2008,Renaud2013,Fujimoto2014a,Dobbs2014}. While major events such as galactic mergers and tidal passages can occur during a galaxy's evolution and produce serious disturbances in the environment \citep{Hopkins2013,Pettitt2017}, the star formation efficiency of a late-type galaxy such as the Milky Way will principally be controlled by its quiescent potential. Numerous studies exist where authors study simulations of clouds formed in disc galaxies \citep{ShettyOstriker2008,Dobbs2012,Benincasa2013,Ward2016}, yet none specifically address the impact of the the variety of rotation curves and mass models observed in external galaxies has on cloud formation and properties. It is expected that the galactic rotation curve will directly impact cloud formation, specifically through the effect of differing shear \citep{Elmegreen1993,Hunter1998,Tan2000}. This should differ depending on the how the rotation changes with radius, e.g. depending on whether the rotation curve is flat, centrally peaked or rising with increasing radius. Observations indicate that shear in galactic discs has somewhat of an effect on global ISM/cloud structure \citep{Seigar2005,Luna2006,Elson2012} and how gas behaves when leaving/entering spiral arms \citep{Koda2009,Miyamoto2014}. Adiitionally, there exists some simulation-based evidence of high shear rates impeding star formation \citep{Weidner2010,Hocuk2011}. However, there is also strong evidence that shear has no impact on cloud/star formation in both the Milky Way \citep{Dib2012} and M51 \citep{Meidt2013}, and that the shape of rotation curves has no impact on global star formation rates \citep{Watson2012}.

Accordingly, we implemented various simulations that produce different galactic background potentials to investigate their effects on the star-forming clouds. The paper is constructed into four parts. The following Section \ref{Numerical methods} presents the code and numerical algorithm used in this analysis as well as the initialization of the simulations. The results of the global ISM and GMCs will be discussed and compared in Section \ref{Results}. We conclude some main points in the final section.

%%%%%%%%%% Numerical methods %%%%%%%%%%
\section{Numerical methods}
\label{Numerical methods}

\subsection{The code}
\label{The code}

To investigate the effects of different galactic environments, we conducted a series of simulations using {\tt ENZO}; a three-dimensional adaptive mesh refinement (AMR) hydrodynamics code \citep{BryanNorman1997, Bryan1999, Bryan2014} previously utilized successfully in galactic-scale simulations by \cite{TaskerTan2009, Fujimoto2014a, Utreras2016, Jin2017}. One of the strengths of this method is the ability to model multiphase gases over a wide range of temperature, density and pressure, allowing a self-consistent ISM to be easily developed. {\tt ENZO} utilises a three-dimensional version of the {\tt ZEUS} hydrodynamics algorithm to evolve the gas with self-gravity and radiative cooling. For all simulations presented here we adopt a quadratic artificial viscosity factor of 2.0. The radiative cooling model for solar metallicity from \citet{Sarazin1987} was used to cool the gas down temperatures of $T=10^4$\,K, after which the gas followed the rates from \cite{Rosen1995} to further cool to the upper end of the atomic cold neutral medium at $T=300$\,K \citep{Wolfire2003}. While the internal temperature of a GMC is around $10$\,K, this higher cut-off allows us to crudely allow for support below our resolution, including internal cloud turbulence and also the presence of magnetic fields. In this study, we omit the star formation and stellar feedback to study the formation and evolution of the gas clouds without internal disruption. 

The galaxies are modelled in a three-dimensional simulation box of size 32\,kpc across with a root grid of $128^3$ cells and additional five levels of refinement, producing a limiting resolution (smallest cell size) of 7.8\,pc.  Refinement was implemented whenever the Jeans length dropped below four cell widths, as suggested by \citet{Truelove1997} to prevent artificial fragmentation. On the finest grid cell level, where further refinement is not possible, a pressure floor was introduced in the form of a polytrope with an adiabatic index of 2.0, to terminate the collapse at a finite density. 

\subsection{The initial conditions}
\label{Initial condition}

In this paper, we present a series of simulations divided into four sets of three (in one case, two) galaxy disc runs. Within each set of simulations, the gas disc sits in a background potential that produces one of three types of rotation curve; a circular velocity that rises with galactocentric radius ({\bf Rise}), a circular velocity that decreases with galactocentric radius ({\bf Decrease}) and flat profile where the circular velocity is approximately constant over the disc ({\bf Flat}). Each of these trends has been observed in disc galaxies, {such as the flat rotation curve of M31, the decreasing curve of NGC 4414 and the rising curve of M81 \citep{Widrow2003, Bosma1998, Feng14}. This ensemble of rotation curves are also utilised in simulations of galaxy interactions by Pettitt et al., (submitted). As we aim to compare the different environments produced by the shape of these rotation curves, we normalise the velocity to give a rotation speed around $200$\,km\,s$^{-1}$ at 10\,kpc, similar to the rotation speed for the Milky Way \citep{Sofue2001}, for all runs except the {\it Extreme} set. The run labels and properties are listed in Table~\ref{run_names}.

\begin{table*}
\begin{center}
\begin{tabular}{lcccccc}
	\multicolumn{2}{c}{Simulation} & Rotation curve & Circular velocity ($\textrm{kms}^{-1}$) & Spiral & Gas profile & Disk gas mass ($10^9 \textrm{M}_\odot$)\\ \hline
	\multirow{3}{*}{Fiducial} & FiR & Rise & \multirow{3}{*}{$\sim 200$} & \multirow{3}{*}{No} & \multirow{3}{*}{Constant} & 2.7 \\
    &FiD & Decrease & & & & 3.4 \\ 
    &FiF & Flat & & & &  3.2  \\ \hline
    
    \multirow{3}{*}{Extreme} &ExR & Rise & \multirow{3}{*}{$\sim 500$ } & \multirow{3}{*}{No} & \multirow{3}{*}{Constant} & 7.1 \\
    &ExD & Decrease & & & &  5.1 \\ \hline

    \multirow{3}{*}{Stellar profile} &StR & Rise & \multirow{3}{*}{$\sim 200$ } & \multirow{3}{*}{No} & \multirow{3}{*}{Variable} & 5.9  \\
    &StD & Decrease & & & & 5.9 \\ 
    &StF & Flat & & & & 5.9 \\ \hline
    
	\multirow{3}{*}{Spiral} &SpR & Rise& \multirow{3}{*}{$\sim 200$} & \multirow{3}{*}{Yes} & \multirow{3}{*}{Constant} & 2.7\\
    &SpD & Decrease & & & &  3.4 \\ 
    &SpF & Flat & & & & 3.0 \\ \hline
\end{tabular} 
\caption{Summary of the simulations presented in this paper. `Variable' implies that the gas disc profile of the {\it St} models is proportional to the stellar density profile and so varies between the rotation curves. Note that even though the gas profile and initial stability of the disc is identical in the non-{\it St} simulations, the dependence on circular velocity (via $\kappa$) causes variations in the gas mass.}
\label{run_names}
\end{center}
\end{table*}

\subsection{The background potential}
\label{ic:potential}

The galaxies are set-up as isolated gas discs in a static background potential that is comprised of a bulge, stellar disc and dark mater halo. This three component form for the potential follows that of \cite{Pichardo2003}, and can be combined to give generally good agreement with Milky Way-like rotation curves \citep{Pettitt2014}. 

The stellar disc component of the potential is described by the standard Miyamoto-Nagai form \citep{MiyamotoNagai1975}:
\begin{equation}
\Phi_d=\frac{GM_d}{(R^2+[a_d+(z^2+b_d^2)^{1/2}]^2)^{1/2}}
\label{Phi_disk}
\end{equation}
where $M_d$ is the stellar disc mass, $x,y,z$ are the position co-ordinates from the galactic centre, which give the square of the galactocentric radius in the plane of the disc, $R^2 = x^2 + y^2$, and $a_d$ and $b_d$ are the radial and vertical scale lengths of the stellar disc, respectively. The associated density profile takes the form:
\begin{equation}
\begin{aligned}
\rho(R,z)=& \frac{b_d^2M_d}{4\pi}\times \\
			   & \frac{a_dR^2+[a_d+3(z^2+b_d^2)^{1/2}][a_d+(z^2+b_d^2)^{1/2}]^2}{\{R^2+[a_d+(z^2+b_d^2)^{1/2}]^2\}^{5/2}(z^2+b_d^2)^{3/2}}.
\label{Phi_diskRho}
\end{aligned}
\end{equation}
which is utilised for the {\it Stellar Profile} calculations.

The form of the bulge potential is taken from \cite{Plummer1911}:
\begin{equation}
\Phi_b=-\frac{GM_b}{\sqrt[]{r^2+r_b^2}}
\label{Phi_bulge}
\end{equation}
where $r_b$ controls the size of the flattened density profile in the bulge core, $M_b$ is the bulge mass and $r^2=x^2+y^2+z^2$ is the spherical radius. 

The spherical dark matter halo is modelled by \cite{AllenSantillan1991}:
\begin{equation}
\begin{aligned}
\Phi_h=&-\frac{GM_h(r)}{r}\\
          &-\frac{GM_{h,0}}{\gamma r_h}\left[-\frac{\gamma}{1+(r/r_h)^\gamma}+\ln\left( 1+\left(\frac{r}{r_h}\right) ^\gamma\right)\right]_r^{r_{h, max}}
\end{aligned}
\end{equation}
where the halo scale length is $r_h$, the halo mass is $M_{\rm h,0}$, the truncation distance is $r_{h,{\rm max}}=100$\,kpc and $\gamma=1.02$. The mass inside a radius $r$ within the halo is given by:
\begin{equation}
M_h(r)=\frac{M_{h,0}(r/r_h)^{\gamma+1}}{1+(r/r_h)^\gamma}.
\end{equation}

Finally, we have an additional time-dependent non-axisymmetric spiral potential used in the {\it Spiral} calculations. This is akin to a density wave spiral \citep{LinShu1964}. The adopted potential is based on the form of \citet{Wada2004, Khoperskov2013} and is included as a perturbation to the potential of the stellar disc:
\begin{equation}
\Phi_{sp}=\frac{(R/a_{sp})^2\cos[2\phi+\Omega_{sp}t-{\rm cot}(\alpha)\ln(R/b_{sp})]}{(1+(R/a_{sp})^2)^{3/2}} 
\label{Sp_eq}
\end{equation}
where $t$ is time, $\phi$ is the cylindrical angle, $a_{sp}= 7.0$\,kpc is the radial scale length of the spiral and $b_{sp} = 3.5$\,kpc is a scale length defining the orientation at $t=0$. $\Omega_{sp} = 20$\,$\textrm{kms}^{-1}\textrm{kpc}^{-1}$ is the pattern speed of the sprial, which is similar to that of the observed spiral arms in the Milky Way \citep{Gerhard2011}. We chose $\alpha = 15^\circ$ for the pitch angle; a value in keeping with observed two-armed spirals \citep{Grosbol2004}. While density waves produced by smooth potentials are a common method of producing spiral arms, simulations using live $N$-body stellar discs tend to show a different type of spiral feature. These spiral arms instead are a strong function of disc stability and galactic mass model, and are seen to be highly dynamic, transient and recurrent in nature (see review of \citealt{Dobbs2014}). The primary difference is that they rotate with the material speed of the disc, rather than at a fixed pattern speed, and so may be expected to differ in their impact on GMC's due to locations of spiral shocks. Investigations into the differences between these two spiral models in respect to the structure of the ISM have been the subject of a select few studies \citep{{Dobbs2010, Grand2015}}. \citet{Baba2016} in particular focused on GMC properties in each spiral model and find differences primarily in how GMC's are destroyed, with a fixed potential-based spirals pulling GMC's apart into spurs as they leave the spiral arm.

The combined galactic potential for each model can be written as \citep{Cox2002}:
\begin{equation}
\Phi=\Phi_d(1+\varepsilon\Phi_{sp})+\Phi_r+\Phi_h
\label{full_potential}
\end{equation}
where $\varepsilon=0.05$ for the {\it Spiral} runs and $\varepsilon=0.0$ for {\it Fiducial}, {\it Extreme}, and {\it Stellar Profile}, and describes the relative amplitude of the spiral stellar density wave.

The component scale lengths and masses are varied to create the {\it Rise}, {\it Decrease}, and {\it Flat} curves, with the chosen values listed in Table~\ref{potential parameters}. We note that while these potentials represent the global velocity field of a galactic system, they lack the details gained by using an $N$-body stellar disc, which is susceptible to time-dependent spiral and bar perturbations that are strongly dependent on the disc--bulge--halo mass ratios and scale lengths.

\begin{table*}
\begin{center}
\begin{tabular}{lccccccc}
\multicolumn{3}{c}{Term \& Description} & FiR & FiD & FiF & ExR & ExD \\ \hline
$M_d$ &[$10^{10}$\,M$_\odot$] & Disc mass	    & 5.99 & 5.99 & 6.85 & 51.36 & 2.57 \\
$M_{h,0}$ &[$10^{10}$\,M$_\odot$] & Halo mass & 12.84 & 8.56 & 10.70 & 42.80 & 5.35\\
$M_b$ &[$10^{10}$\,M$_\odot$]& Bulge mass     & 0.70 & 1.40 & 1.40 & 2.80 & 11.20\\
$a_d$ &[kpc] & Disc radial scale length      & 5.30 & 1.86 & 3.45 & 5.30 & 10.60 \\
$b_d$ &[kpc] & Disc vertical scale length    & 2.50 & 0.87 & 1.63 & 0.25 & 0.25 \\
$r_h$ &[kpc] & Halo radial scale length      & 12.20 & 24.00 & 14.64 & 12.20 & 36.00\\
$r_b$ &[kpc] & Bulge radial scale length     & 1.95 & 0.78 & 1.05 & 2.34 & 0.59 \\ \hline
\end{tabular} 
\caption{Galactic axisymmetric potential parameters used to produce the different rotation curves.}
\label{potential parameters}
\end{center}
\end{table*}

\subsection{The gas disc}
\label{ic:gasdisc}

In the {\it Fiducial}, {\it Extreme} and {\it Spiral} run sets (see Table~\ref{run_names}), the initial density profile for the gas disc is the same as that presented in \citet{TaskerTan2009}. In brief, the radial profile of the gas is derived assuming a constant value for the Toomre $Q$ parameter for gravitational instability \citep{Toomre1964}:
\begin{equation}
Q=\frac{\kappa c_s}{\pi G\Sigma_g}
\label{Q_eq}
\end{equation}
where $c_s$ is the gas sound speed,  $\Sigma_g$ is the gas surface density, and $\kappa$ is the epicycle frequency, defined by:
\begin{equation}
\kappa^2 = 4\Omega^2 + R \frac{{\rm d}\Omega^2}{{\rm d}R}
\end{equation}
where $\Omega$ is the rotational frequency of the galaxy. In the original 2D analysis, \citet{Toomre1964} calculated that discs with $Q < 1$ would be unstable to gravitational fragmentation. This calculation was repeated by \citet{Goldreich1965} for 3D discs and lowered slightly to $Q < 0.67$. We adopt a vertical density profile proportional to $\textrm{sech}^2(z/z_h)$, where $z$ is the vertical coordinate from the disc mid-plane and $z_h$ is the vertical scale height, based on observations of HI in the Milky Way \citep{Binney1998}. We use $z_h=290$\,pc in our simulations, corresponding to the value at approximately the Solar radius. 
The disc gas is initialised according to Equation \ref{Q_eq} by using a disc with a gas density profile given by $\Sigma_g=\int^\infty_{-\infty}\rho_0$\,sech$^2(z/z_h)dz=2\rho_0z_h$, where $\rho_0$ is the mid-plane ($z=0$) density. The complete gas distribution then becomes:
\begin{equation}
\rho(R,z)=\frac{\kappa c_s}{2\pi GQz_h} \textrm{sech}^2\left(\frac{z}{z_h}\right).
\end{equation}
For the main section of the disc between radii $2 < R < 10$\,kpc, the initial gas surface density for each calculation is set by fixing a value of $Q=3$, and an initial sound speed of $c_s = 9$\,km\,s$^{-1}$. As the gas cools, $Q$ drops below the threshold for gravitational fragmentation. For the inner and outermost regions, $Q = 20$ and the low density gas remains stable to gravitational collapse.

It is worth noting that a $Q$ parameter also exists for the stars, being a function of velocity dispersion in the stars as opposed to sound speed. A low $Q$ in the stars would result in instabilities and fragmentation of the stellar disc (e.g. \citealt{1985MNRAS.217..127S,2011ApJ...730..109F}). As our stellar component here is a simple static potential, it effectively assumes a high $Q$ in the stars.

In the {\it Stellar Profile} runs we instead set gas up following the stellar density profile of Equation~\ref{Phi_diskRho}. The parameters $a_d$ and $b_d$ and were set the same as the background potential, with a mass scaling factor added so that the disc gas mass is approximately $\sim 10\%$ of stellar disc mass, in accordance with the gas fractions measured in observed galaxies \citep[e.g.][]{Kalberla2007}. The exception is the slightly lower fraction for the {\it Stellar Flat, StF} disc, in order to have the same total gas mass in all three cases.

\subsection{Differences between the runs}

\begin{figure*}
\centering
	\includegraphics[width=13.0cm]{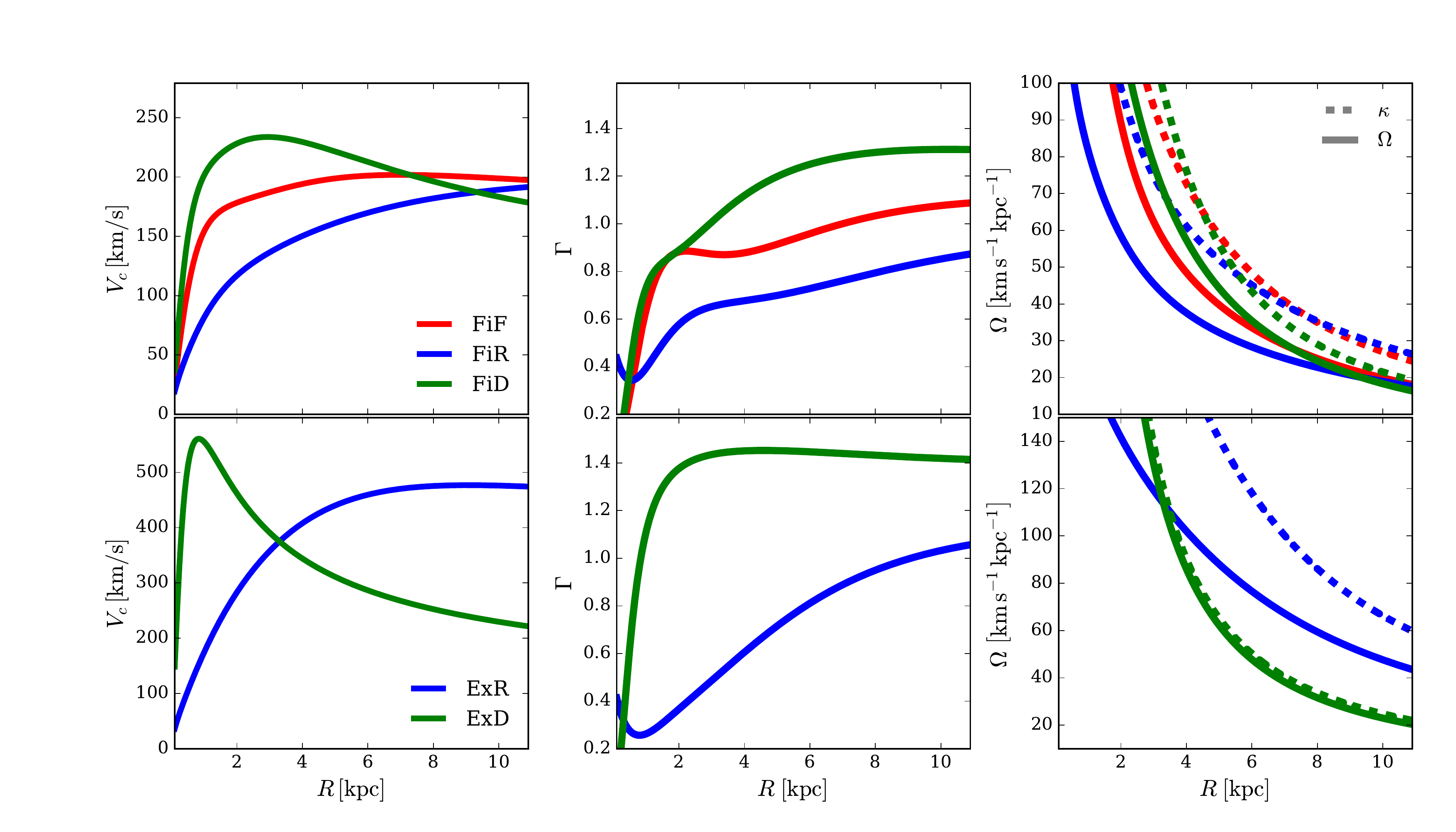}
	\caption{The circular velocity (left), shear (middle) and rotational frequencies (right) for the different simulations. Top row shows the profiles for the {\it Fiducial} runs, while the lower row shows the {\it Extreme} set. The {\it Stellar Profile} and {\it Spiral} runs have the same curves as for the {\it Fiducial} runs. The right-hand plots show both the circular frequency, $\Omega$ (solid lines) and the epicycle frequency, $\kappa$ (dashed lines).}
	\label{circular_velocity}
\end{figure*}

In the {\bf Fiducial} (Fi) runs, the isolated gas disc and background potential are set up as described above, with three identical discs following the {\it Rise}, {\it Decrease} and {\it Flat} rotation curves with a circular velocity at the outer edge of around 200\,km\,s$^{-1}$. For the {\bf Extreme} (Ex) runs, the circular velocity is dramatically increased to around 500\,km\,s$^{-1}$. These velocities are intentionally high to exemplify the impact of the different rotational curves. That said, such high rotation rates are not unheard of in observed galaxies, such as NGC 2699 \citep{Noordermeer2007}. Since models for the {\it Flat} and {\it Rise} rotation curves for the {\it Extreme} runs look very similar, we only present {\it Rise} and {\it Decrease} for this set of simulations. The gas disc for these runs remains the same as in the {\it Fiducial} cases.

The rotation curves, associated shear rates and rotation frequencies are shown in Figure~\ref{circular_velocity} for the {\it Fiducial} runs (top row) and the {\it Extreme} runs (bottom). The shear rate is defined as: 
\begin{equation}
\Gamma=1-\frac{R}{V_c}\frac{dV_c}{dR}, 
\end{equation}
where $V_c$ is circular velocity at galactocentric radial distance $R$, and $\Gamma$ relates to the Oort constant $A$ via $A=\Gamma\Omega/2$. The right columns show both the circular ($\Omega$) and epicycle ($\kappa$) frequencies. The top row shows the $V_c$, $\Gamma$, $\Omega$ and $\kappa$ for our {\it Fiducial} runs, with circular velocities around 200\,km\,s$^{-1}$, while the bottom row showns the {\it Extreme} runs whose circular velocity reaches to around 500\,km\,s$^{-1}$. 

Although the {\it Rise} and {\it Decrease} rotation curves have larger gradients (in opposing directions) than the {\it Flat} curve, the shapes of shear rate along the galactocentric radius shows a similar trend. Shear rates in {\it Flat} and {\it Decrease} models show a considerable drop near the  centre ($R < 2.5$\,kpc) and all three gradually rise from the mid-disc outward. The {\it Decrease} curve consistently has the highest shear rate (about 1.3 at the disc edge), followed by the {\it Flat} (1.1) and {\it Rise} (0.8). The shear rate in spiral galaxies is believed to correlate with the pitch angle of the arms \citep{Grand2013,Seigar2005b,Seigar2006}, with stronger shear leading to tightly wound spiral arms.  We would naively expect the shear to play a direct role in the formation of GMCs, with the highest shear rates suppressing cloud formation.

In the {\it Extreme} simulation set, the {\it Decrease} run shows a much greater difference compared to the {\it Fiducial} counterpart, having greater circular velocity and shear in the inner part of the disc. The shear then remains constantly large from $R > 2$\,kpc. This produces a very different environment to the {\it Extreme Rise} run, whose shear shows less difference from the {\it Fiducial} case. At the disc edge, both the {\it Fiducial} and {\it Extreme} complementary runs have similar shear values.

The orbital frequencies show similar trends across all {\it Fiducial} cases, with increasing values for the circular frequency, $\Omega$, from the {\it Rise} to {\it Flat} then {\it Decrease} runs. The epicycle frequencies play a role in the stability of the disc to collapse (see Equation\;\ref{Q_eq}). Each model in the {\it Fiducial} set only has slight differences in $\kappa$, with each decreasing in roughly the same gradient, and all being roughly equivalent at around 5\,kpc. In the {\it Extreme} runs, however, the {\it Extreme Rise} curve has a much higher value for $\kappa$ compared with all other runs and its own $\Omega$, which can potentially help stabilise the disc. The {\it Extreme Decrease} meanwhile, have $\Omega$ and $\kappa$ coinciding with one another. 

Overall, the {\it Rise} curves for both the {\it Fiducial} and {\it Extreme} cases are characterised by small shear and a steady rise in circular velocity to create a system with solid body--like rotation. Conversely, the {\it Decrease} curves are characterised by a much higher shear rate to give Keplerian-like rotation.

Our third set of runs, {\bf Stellar Profile} (St), changes the gas profile of the {\it Fiducial} case to a profile that follows the stellar disc distribution (Eq.~\ref{Phi_diskRho}). As such, each different rotation curve model will have a different gas disc, providing a more realistic gas surface density profile at the expense of being able to directly compare the isolated impact of differing the rotation curve (as $\Sigma_g(R)$ is now not the same between rotation models). 

Our final set of calculations, {\bf Spiral} (Sp), include the two-armed spiral perturbation of Equation~\ref{Sp_eq} that moves with a fixed pattern speed. The resulting resonances from the spiral perturbation are shown in Figure~\ref{resonances} for each of the {\it Spiral} runs. The blue horizontal line shows the spiral pattern speed, and the green vertical lines show where this intersects the various frequencies inherent to the axisymmetric potential. While the co-rotation radius (solid vertical line) is approximately the same for each disc, the Lindblad Resonances ($\pm\kappa/2$) and 4:1 resonance radii ($\pm\kappa/4$) are quite different in each case, implying we would expect to see a different response to the spiral in each model (e.g. \citealt{Elmegreen1989}). While there are other ways of producing spiral structures in simulations of disc galaxies, we chose a density wave-like spiral perturbation due to the well defined nature of such a static potential, and the difficulty in producing long lived grand design spirals in simulations by other methods such as swing amplified instabilities or tidal interactions (see \citealt{Sellwood2011,Dobbs2014}).

\begin{figure}
\centering
	\includegraphics[width=8.5cm]{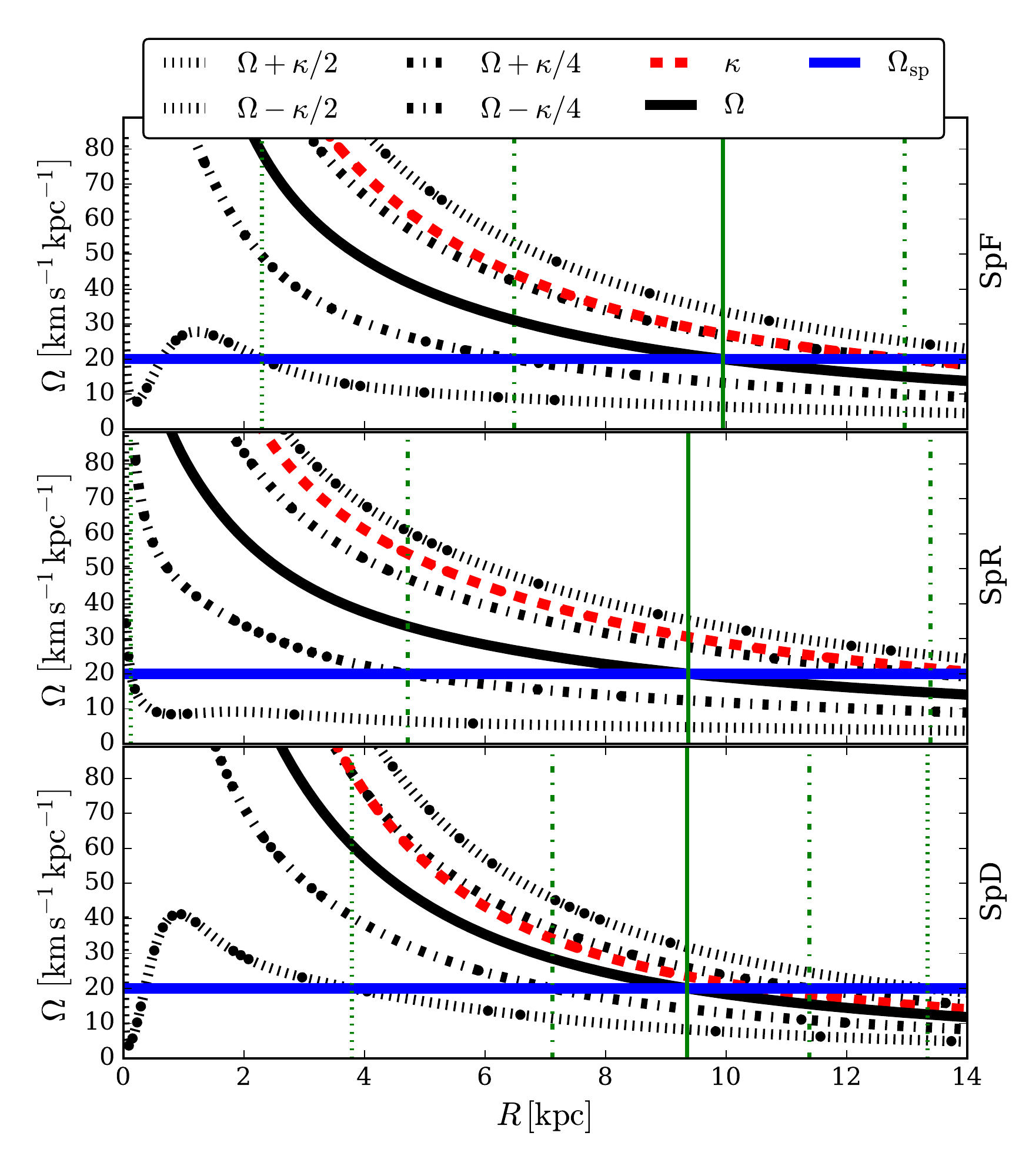}
	\caption{The multiple rotation frequencies within the different discs with an imposed spiral perturbation. Each panel shows the various combinations of $\Omega$ and $\kappa$, with the blue horizontal line showing the spiral rotation frequency. The green lines indicate the locations of various resonances (solid is co-rotation, dotted are I/OLR and dot-dashed are the 4:1 ``ultraharmonic" resonances).}
	\label{resonances}
\end{figure}

The background potential for {\it Stellar Profile} and {\it Spiral} remains the same as those in the {\it Fiducial} runs. Thus, so does the rotation curve, shear, and intrinsic orbital frequencies.

\subsection{Cloud identification}
Potential star-forming clouds were identified within the main disc region between $2.5 < R < 8.5$\,kpc, away from the boundary with the low density areas. Clouds are identified as continuous structures with a gas density over $n_{\rm thresh} \ge 100$\,cm$^{-3}$, found via the contouring algorithm in the analysis software, {\tt yt} \citep{Turk2011}. This identification density threshold agrees with the average density of observed galactic GMCs. We do not follow the production of molecular gas, but assume that gas above this density would consist of a molecular core surrounded by an atomic envelope of hydrogen. The computational cells that lie within the cloud contour are used to calculate the cloud bulk properties in the same method as in \citet{Fujimoto2014a}.

%%%%%%%%%% Results %%%%%%%%%%
\section{Results}
\label{Results}
%%%%%%%%%% subsection %%%%%%%%%%
\subsection{The interstellar medium}
In this section we discuss the global changes in the ISM between our calculations. Comparisons are conducted at $t=300$\,Myr, when there is relatively little in change in global disc structure. \citet{TaskerTan2009} observed that although the disc fully fragments after $\sim 140$\,Myr, the rate of cloud formation and destruction is nearly constant from 200\,Myr to 300\,Myr.

%%%%%%%%%%% subsubsection %%%%%%%%%%%
\subsubsection{Disc structure}
\label{The gas disc}

\begin{figure*}
\centering
	\includegraphics[width=\textwidth , trim = 0mm 15mm 0mm 0mm]{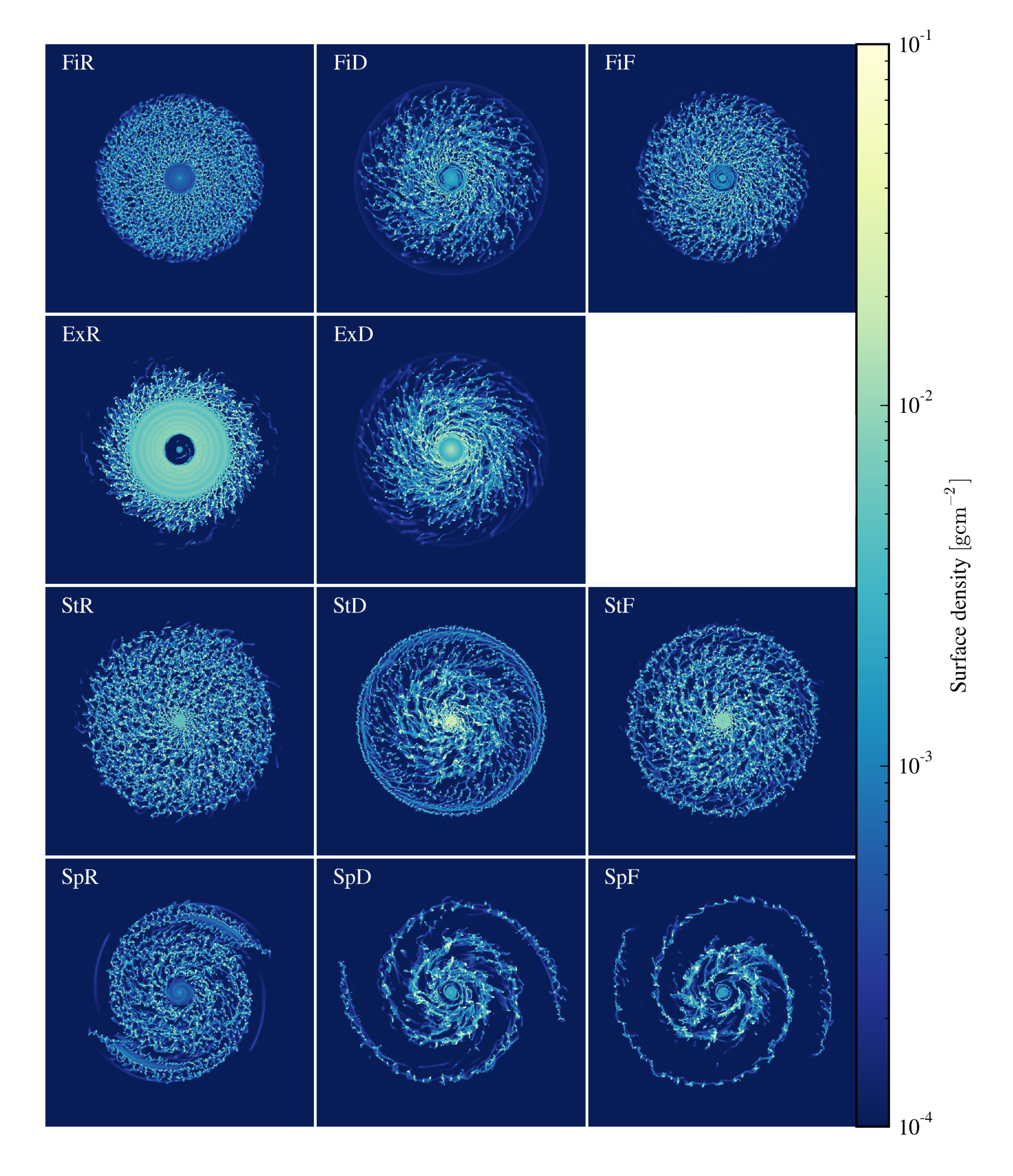}
	\caption{The structure of each gas disc after $t = 300$ Myr of evolution. Projections show a $32\times 32$\;kpc region and show the gas surface density integrated vertically over $|z|\le 1$ kpc. Each row shows a different model type (Fi, Ex, St, Sp) and each column the different type of rotation curve ({\it Rise}, {\it Decrease} and {\it Flat}). The gas discs are fully fragmented into clouds in most of models at all radii except for the ExR model.} 
	\label{projection_300}
\end{figure*}

Figure~\ref{projection_300} shows the surface density of gas discs in a $32\times 32$\,kpc projection at $t=300$ Myr. The gas surface density, $\Sigma_g$, is integrated vertically over $|z| \le 1$ kpc above and below the disc. To quantify what is driving stability in each model, we calculate the Toomre $Q$ (Eq.\;\ref{Q_eq}) in all of our simulations. However, $Q$ contains no information about the shear in each disc (i.e. there is no $\Gamma$ dependance). The stability of material to shear collapse has been parameterized by \citet{Dib2012}, based on the work of \citet{Elmegreen1993,Hunter1998}, into the parameter $S$ given by:
\begin{equation}
S=\frac{2.5 \sigma_g A}{\pi G\Sigma_g}
\label{S_eq}
\end{equation}
where $A$ is an Oort constant. This is related to our definition of shear rate by $A=\Gamma\Omega/2$. If $S>1$ then shear will succeed in impeding structure growth, while $S<1$ would imply shear is ineffective. Similar definitions of shear support have been used by \citet{Seigar2005,Luna2006,Meidt2013}. Figures~\ref{Qimage} and \ref{Simage} show the $Q$ and $S$ parameters respectively as a function of galactocentric radius. In Table\;\ref{cloudNo} we show the number of clouds within each simulation at 300\,Myr, though our discussion of cloud properties will be the topic of Section~\ref{Cloud_properties}.

\begin{table}
\begin{center}
\begin{tabular}{lc}
Calculation & Cloud number\\ 
\hline
\hline
FiR & 1897 \\ 
FiD & 1029 \\ 
FiF & 1389 \\ 
\hline
ExR & 945 \\ 
ExD & 966 \\ 
\hline
StR & 1464 \\ 
StD & 697 \\ 
StF & 1134 \\ 
\hline
SpR & 979 \\ 
SpD& 334 \\ 
SpF & 335 \\ 
\hline
\end{tabular} 
\caption{Cloud numbers for all simulations at a time of 300\,Myr.}
\label{cloudNo}
\end{center}
\end{table}

The top row of Figure~\ref{projection_300} shows our {\it Fiducial} runs; left is the {\it Rise} rotation curve (FiR), followed by {\it Decrease} (FiD) and then {\it Flat} (FiF) rotation curves. FiR shows a uniform clumpy structure (containing 1897 clouds), and FiD shows the least clumpy structure (with only 1029 clouds). FiF is between the two (1389 clouds), which is to be naively expected from the shear, $\Gamma$, being between FiR and FiD at all radii.

Inspection of $Q$ and $S$ show that FiR is unstable to shear and borderline unstable to Toomre (pressure+Coriolis+gravitational) collapse, with $Q$ lying right on the 3D stability limit in Figure~\ref{Qimage}. The lack of any strong change in $Q$ and $S$ as a function of radius leads to a uniformly fragmented disc, which is to be expected from a solid body-like rotation curve.

FiD and FiF have greater rates of shear than FiR (Figure\;\ref{circular_velocity}) which is the likely cause as to why they have formed fewer clouds than FiR. In particular the Keplerian-like rotation model FiD appears shear stable for $R>6.5$\,kpc, even though the gas appears $Q$ unstable at all radii (Figure\;\ref{Simage}). This is the cause of the dearth of collapse and cloud formation in the outer regions of FiD in comparison to FiF and FiR, which show a more uniform collapse.

\begin{figure}
\centering
\includegraphics[width=8.5cm]{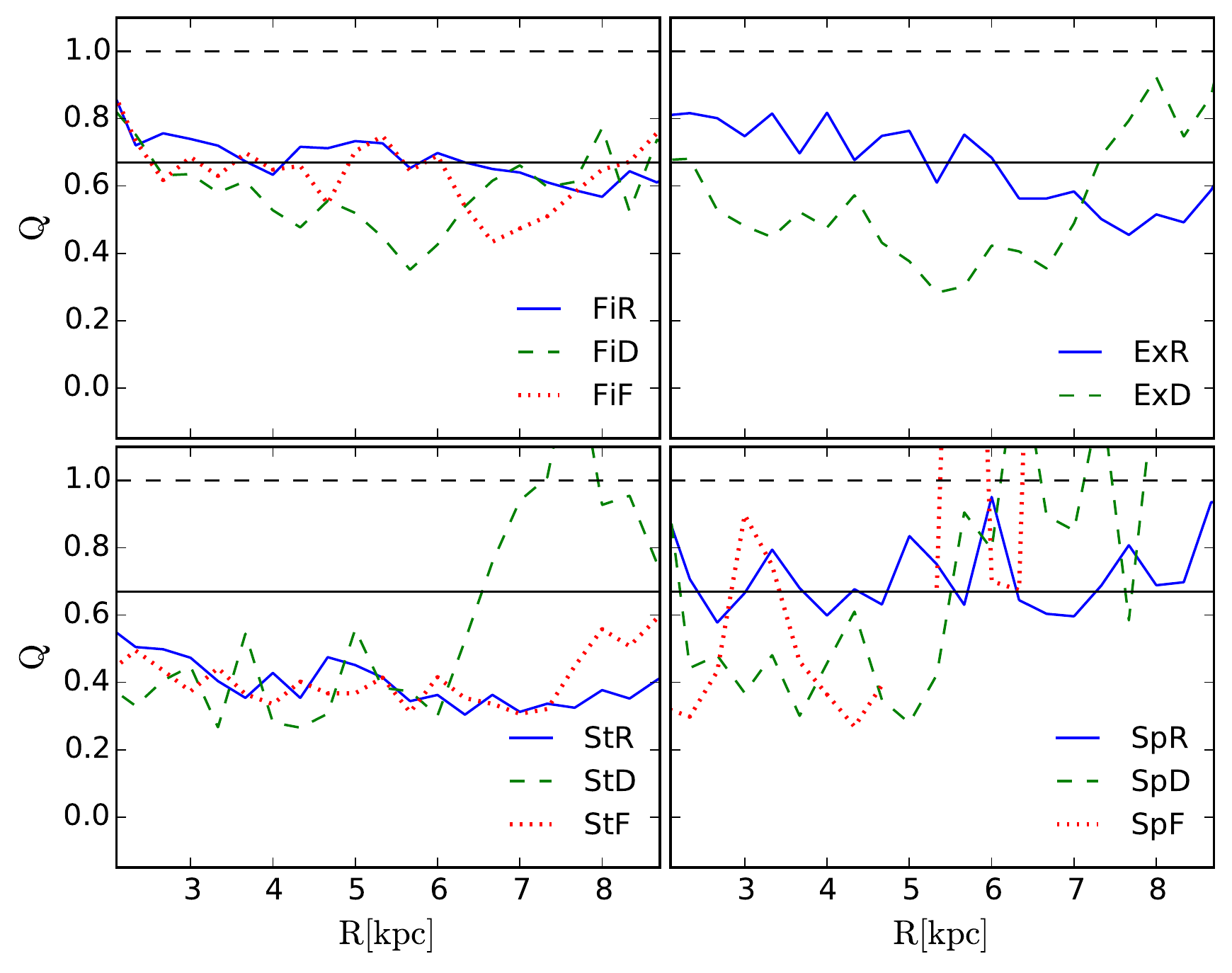}
\caption{The Toomre-$Q$ parameter in the gas for all simulations in this study as a function of radius (at $t=300$\,Myr). The horizontal dashed line shows the 2D stability limit of $Q=1$ and the horizontal solid line the approximate 3D stability limit of $Q=0.67$ \citep{Goldreich1965}.}
\label{Qimage}
\end{figure}

\begin{figure}
\centering
	\includegraphics[width=8.5cm]{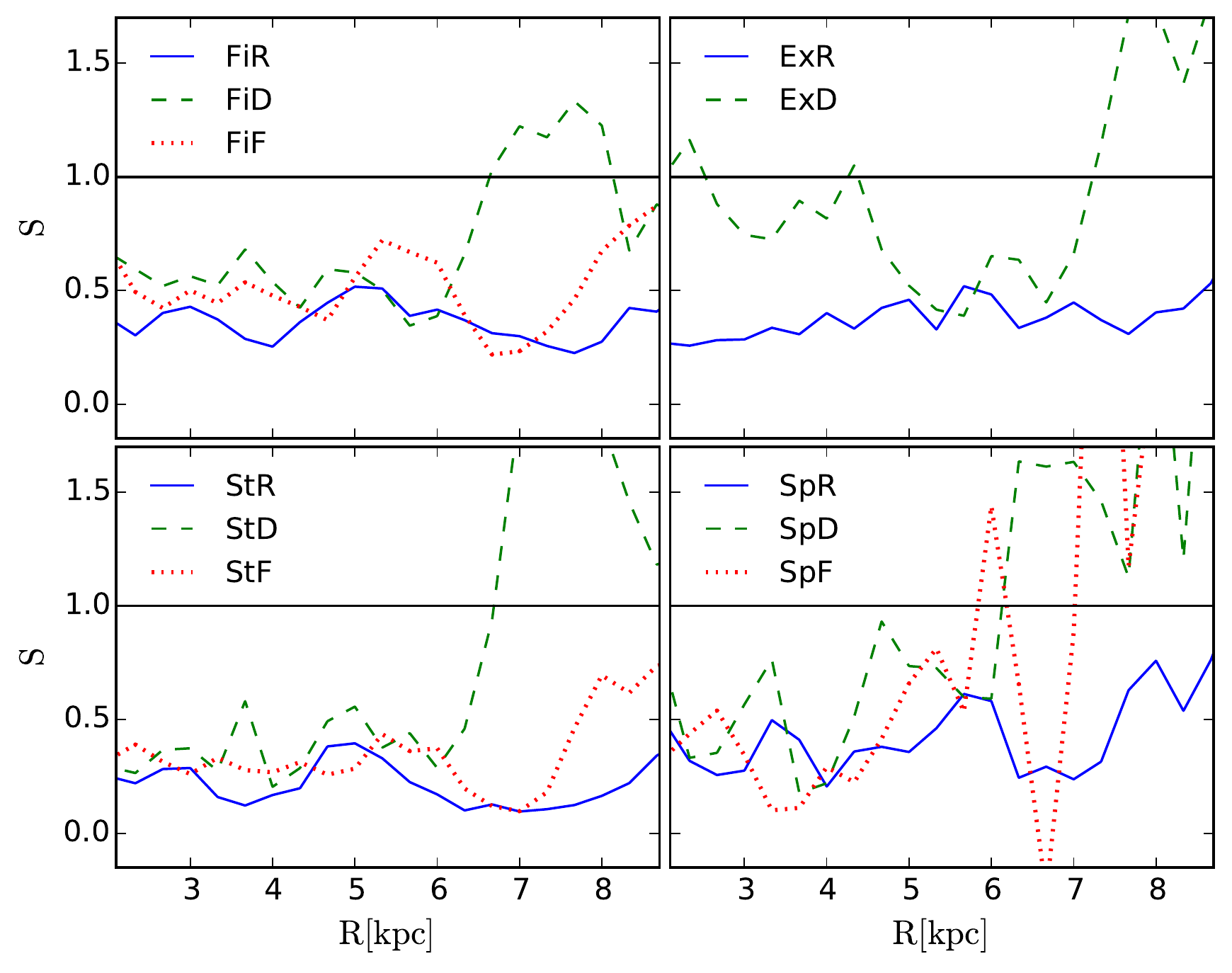}
	\caption{The shear parameter, $S$, in the gas for all simulations in this study as a function of radius (at $t=300$\,Myr). The shear stability limit of $S=1.$ is shown by the solid horizontal line.}
	\label{Simage}
\end{figure}

The second row shows the the two {\it Extreme} curves, the {\it Rising} (ExR, left) and {\it Decreasing} (ExD, centre), where the magnitude of rotation reaches as high as 500km/s. ExR appears to fragment outside-in despite dense gas near the centre, while ExD has a large degree of collapse in the centre compared to the outskirts, producing larger clumps. Surprisingly, the quantity of clouds are nearly identical: ExR has 945 clouds and ExD has 966, though obviously with opposite locations in the disc.

Although $\Gamma$ and $S$ are low throughout the ExR model, thus favouring collapse, there is a clear resistance to collapse in the inner disc. The ExR disc is, however, stable to Toomre collapse in the inner disc, with $Q$ rising above the 0.67 limit for stability in 3D discs. Figure~\ref{circular_velocity} indicates that this is caused by a rather high value of epicyclic frequency, much greater compared to the disc angular frequency than the flat rotation curve relation: $\kappa=\sqrt{2}\Omega$. This high frequency of radial oscillation is enough to provide support against collapse in the inner regions of the disc, despite the very low shear rate. Contrary to this, the ExD disc has the largest value of shear rate of all our calculations. This results in a rise of $S$ at the disc edge similar to FiD, though as $Q$ is somewhat lower than FiD the degree of structure is different compared to the FiD disc, with larger and more massive clumps in ExD.

The third row shows the non-constant gas disc, i.e. where the gas profile traces the stellar surface density of the background potential. The StR, StD and StF runs show the same general trends as the {\it Fiducial} runs. StR is the clumpiest with 1464 clouds, then StF 1134 clouds, the least is StD with 697 clouds. The degree of fragmentation is greater than the {\it Fiducial} case due to a greater gas reservoir (about $6 \times 10^9 \textrm{M}_\odot$, which is approximately twice as large as those of {\it Fiducial} runs). In general they all have lower values of $Q$ than their {\it Fiducial} counterparts in the mid and inner disc due to this increased surface density. The projection of StD shows that it is resisting collapse in the outer disc. This is partially due to a stability to both shear and Toomre collapse, with $Q$ and $S$ being above limiting values in the outer disc. While the disc does have a relatively high shear rate, $\Gamma$, it also has a low gas surface density in the outer disc, as now the gas follows the stellar disc profile which is more centrally concentrated than the StR and StF models.

The last row of Figure \ref{projection_300} shows the {\it Spiral} perturbation runs (SpR, SpD, SpF). They have the same trend in cloud numbers as the {\it Fiducial} and {\it Stellar Profile} runs (SpR 979 clouds, SpF 335 clouds, SpD 334 clouds). These cloud numbers are all systematically lower than the {\it Fiducial} runs, likely due to the lack of a population of interarm clouds. In all cases a 2-armed spiral in the gas can be seen, though each disc has a subtly different response. Gas is clearly swept up into the grand design spiral where clouds are formed, but there is almost no interam population in the SpD and SpF cases. Clouds are thus either being destroyed as the leave the spiral by the rapid change in shear (e.g. \citealt{Miyamoto2014}) or very rapidly pass between arms. While the cloud number is lower, the clouds formed are larger than the axisymmetric cases (see Sec. \ref{Cloud_properties} for a discussion). The SpF model in particular forms regularly spaced out clumps across the spiral arms, similar to the ``beads on a string'' seen in observed \citep{Elmegreen1983,LaVigne2006} and other simulated galaxies \citep{Shetty2006,Renaud2013}.
 
While SpD and SpF show a strong response to the spiral potential, the SpR disc has much weaker gaseous spirals, only showing clear spirality in the outer disc. There is still an effect on the inner disc, which can be seen by comparing the FiR and SpR runs, which only differ by the addition of a spiral potential. To investigate the differing spiral response we plot the {\it Spiral} models in Figure~\ref{spiral_circ} with overlaid radii of various resonance locations highlighted in Figure~\ref{resonances}. In all cases, the gas responds to the spiral in the region between the inner and outer Lindblad resonances (ILR, OLR). In the SpD run the spirals are appearing to dissipate and wind up as they reach the OLR, and in the inner region of the disc inside of the ILR there is no spiral response as expected. This is also true of the SpF model, though the gas disc is truncated before it reaches the OLR region. In all models there is a strong spiral response at the co-rotation radius (CR, the solid red line). 
The SpR calculation has a very limited response inside of the CR, despite the entirety of the inner disk being outside the ILR (which is effectively at $R=0$kpc). The main reason for this is that the mass distribution of the rise models is dark matter dominated. As the spiral potential is set to be a 5\% perturbation of the stellar disc (Eq. \ref{full_potential}) the rising models with their relatively low disc mass will have a weaker spiral structure.

The impact of the spiral's structural rearrangement on the disc stability is indicated in Figures~\ref{Qimage} and \ref{Simage}. Disc SpR unsurprisingly shows the least deviation from the fiducial case, as the disc responds only weakly to the spiral. A minor increase in of $Q$ in the outer disc highlights where the spiral response is the strongest, with the SpR disc also have a reduced cloud spacing by-eye throughout. This implies that while the density wave is too weak to excite a clear gaseous spiral, it still increases the degree of fragmentation, also seen in the value of $Q$ having clear oscillations about the $Q=0.67$ compared to the FiR case. The SpD and SpF show a similar trend, with on average a rise in $Q$ in the outer disc and drop in the inner disc. The drop appears due to the spiral triggering collapse of the high surface density gas near the ILR, which can be seen to create a population of high density cloud complexes compared to their {\it Fiducial} counterparts. The higher Q in the outer regions is likely caused by the strong dearth of inter-arm clouds, with only a small population of dense cloud structures along the arms. This reduces the azimuthally averaged value of $\Sigma_g$ compared to the $Fiducial$ cases, thus causing an increase in $Q$. It is notable that in the case where an external spiral wave has been applied, we would expect gas to fragment along the spiral shock front. This makes $Q$ a somewhat less relevant indicator of the stability of gas to collapse, with a more local $Q$ map as a function of radius and azimuth a better indicator of the stability of specific regions (e.g. \citealt{Miyamoto2014,Goldbaum2015}), though this is beyond the scope of this work.

The shear in all spiral models is, on average, the same between the $Fiducial$ and $Spiral$ runs (as both have the same rotation curves), with the $Spiral$ runs simply having a greater scatter caused by the high arm-interarm contrast impacting $\Sigma_g$ in Eq.\,\ref{S_eq}.

\begin{figure}
\centering
	\includegraphics[width=7.8cm]{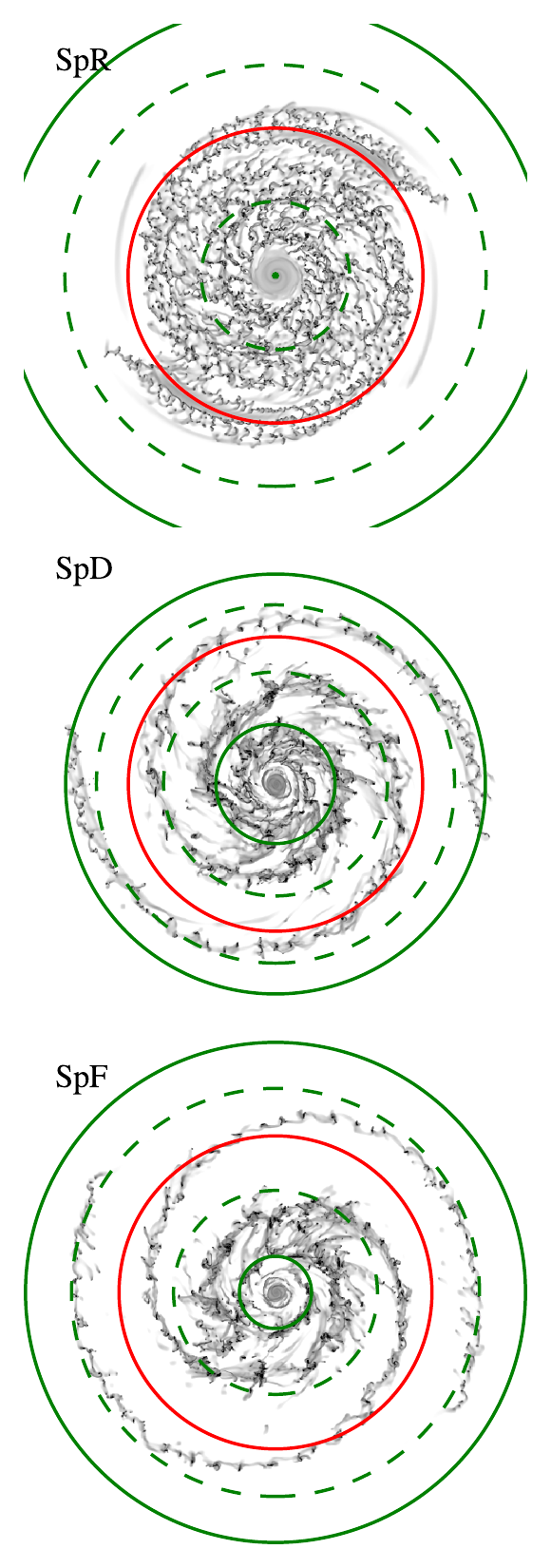}
	\caption{The three spiral models shown in this study after 300\,Myr of evolution. Overplotted are circles illustrating the locations of the ILR (first green solid), inner 4:1 (first green dashed), co-rotation (red solid), outer 4:1 (second green dashed) and OLR (second solid green).}.
	\label{spiral_circ}
\end{figure}

%%%%%%%%%%% subsubsection %%%%%%%%%%%
\subsubsection{Phase plots}

\begin{figure*}
\centering
	\includegraphics[width=\textwidth]{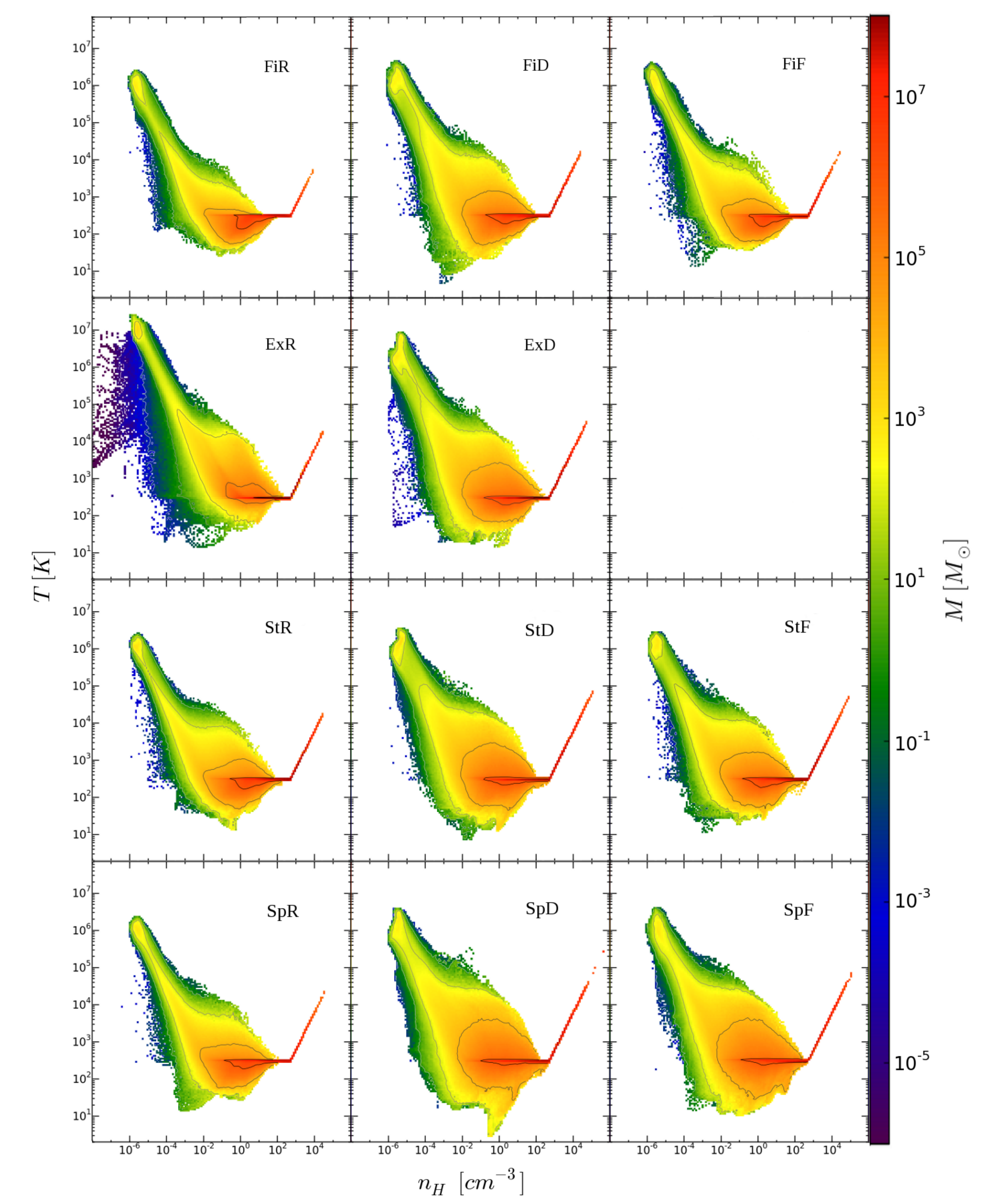}
	\caption{Phase diagrams of temperature and density of ISM gas in each calculation for an annulus of 2.5\;kpc\;$< R <$\;8.5\;kpc and $|z|\leq$\;1\;kpc at a time of $t = 300$ Myr. Each point in $T-n_H$ space is coloured by the mass included in the respective bin.}
	\label{phase_plots}
\end{figure*}

The structure of the ISM can be analysed more quantitively by looking at the gas mass in the two-dimensional density-temperature phase space in Figure~\ref{phase_plots}. The plots show the gas in the galactic discs over an annulus of 2.5 kpc $< R <$ 8.5\,kpc and $|z|\leq$\,1 kpc. The sharp linear trend of high density gas on the right of each plot corresponds to our pressure floor described in Section~\ref{Numerical methods}. This appears to move the densest parts of the cloud gas into the warm ISM ($10^3$\,K $< T < 10^5$\,K), whereas in reality, we expect this to be cold gas. In the discussion below we refer to a hot ISM (HISM) defined as $T > 10^5$ K, a warm ISM (WISM) defined as $10^3$ K $< T < 10^5$ K, and a cold ISM (CISM) defined as $T < 10^3$ K.

Although we discuss three ISM phases, the gas does not form discrete phases but a continuous structure in rough pressure equilibrium, although with a spread of pressures. In all discs, more than 99\% of the gas mass is concentrated in the cold ISM and cloud material with densities running from $n_H > 10^{-4}$. Almost no gas is present in the HISM, which is unsurprising due to the lack of stellar feedback which can generate dense hot gas \citep{Tasker2015}.

The first row of Figure \ref{phase_plots} shows the {\it Fiducial} runs in order of rising, decreasing, and flat rotation curves (FiR, FiD, FiF ) from left to right. Although the surface density images in Figure~\ref{projection_300} and cloud numbers in Table~\ref{cloudNo} show the highest degree of fragmentation occurs in the FiR run, the mass of gas in clouds with $n_H > 100$\,cm$^{-3}$ is less than in the FiD and FiF cases. This suggests that FiD and FiF may be forming a smaller number of more massive clouds compared to the rising rotation curve counterpart. This is consistent with Figure~\ref{Simage}, which showed runs FiD and FiF had a greater support from shear, suppressing the formation of smaller clumps. At lower densities to the clouds, but still within the CISM, FiR has a slightly greater gas mass (0.2\% - 1\%) than the FiD and FiF runs, reflecting the uniform agglomeration within the solid body rotation curve and steady borderline $Q$ value shown in Figure~\ref{Qimage}. With very little gas mass in the HISM, the decrease in CISM in the FiD and FiF runs compared to FiR leads to an increase in the mass in the WISM. This is especially true in the FiD simulation, where shear is providing support against collapse in the outer regions, leading to an increase in WISM in the voids. 

The second row of Figure~\ref{phase_plots} shows the ISM in the {\it Extreme} rotation curve runs. In both cases, there is less gas in the CISM and more in the WISM compared to the {\it Fiducial} runs, although a higher mass of the densest cloud gas. This implies that the gas is more stable to collapse, suppressing smaller structures but allowing the largest clouds to form, in agreement with the previous section which saw high values for the epicycle frequency and shear provide additional support in these discs. The unfragmented inner region of the ExR simulation consists primarily of lower density CISM gas, giving more gas mass at the lower end of our cooling curve at around $n_H \sim 10$\,cm$^{-1}$ than in other runs. The increase in the WISM in the ExD run can be explained by a more extreme version of FiD, whereby collapse is concentrated in the centre of the disc to produce large clouds and a partly-supported, porous structure in the outer region leads to warm gas in the voids. This is visually clear in Figure~\ref{projection_300}.

The {\it Stellar Profile} discs are shown in the third row of Figure~\ref{phase_plots}. The higher total gas mass in these simulations increases the total quantity of mass in clouds. However, despite different gas profiles and total masses, the trends remains similar to the {\it Fiducial} case. The densest gas is present in StD, indicative again of larger clumps being formed. StR has the highest fraction of CISM gas as FiR is to FiF and FiD, and similarly has a uniformly fragmented structure of smaller clumps as shown in Figure~\ref{projection_300}. Discs StD and StF have most gas in the WISM, especially StD which maintains less fragmentation in the outer disc. 

The bottom row shows the discs where a spiral perturbation has been added. Compared to the {\it Fiducial} case, the addition of a spiral potential increases the dense cloud gas and also the mass in the upper part of the WISM around $T \sim 10^4 - 10^5$\,K. This suggests the CISM is being compressed into clouds within the spiral with the lower density WISM sitting between the arms. As the overall cloud number drops significantly as shown in Table~\ref{cloudNo}, it is reasonable to suppose that large clouds are being formed in the spiral arms. This will be shown to be true when we consider cloud properties in Section~\ref{Cloud_properties}. Run SpD shows the most substantial increase in WISM material, followed by SpF. This agrees with the images in Figure~\ref{projection_300}, where those two runs are shown to respond much more strongly to the spiral perturbation compared with SpR, allowing a broader inter-arm region. 

Overall, the simulations with the {\it Decrease} rotation curve (FiD, ExD, StD and SpD) appear to have both the most mass in the densest cloud material and the lower density WISM at the cost of smaller clouds harbouring colder CISM gas. Conversely, the uniform fragmentation of the {\it Rise} curves results in fewer large clumps and a larger CISM at lower density than the clouds. The additional variations between the four simulation sets exacerbate these features, but do not change them.

%%%%%%%%%%% subsubsection %%%%%%%%%%%
\subsubsection{Probability distribution functions}

\begin{figure}
\centering
	\includegraphics[width=8.5cm]{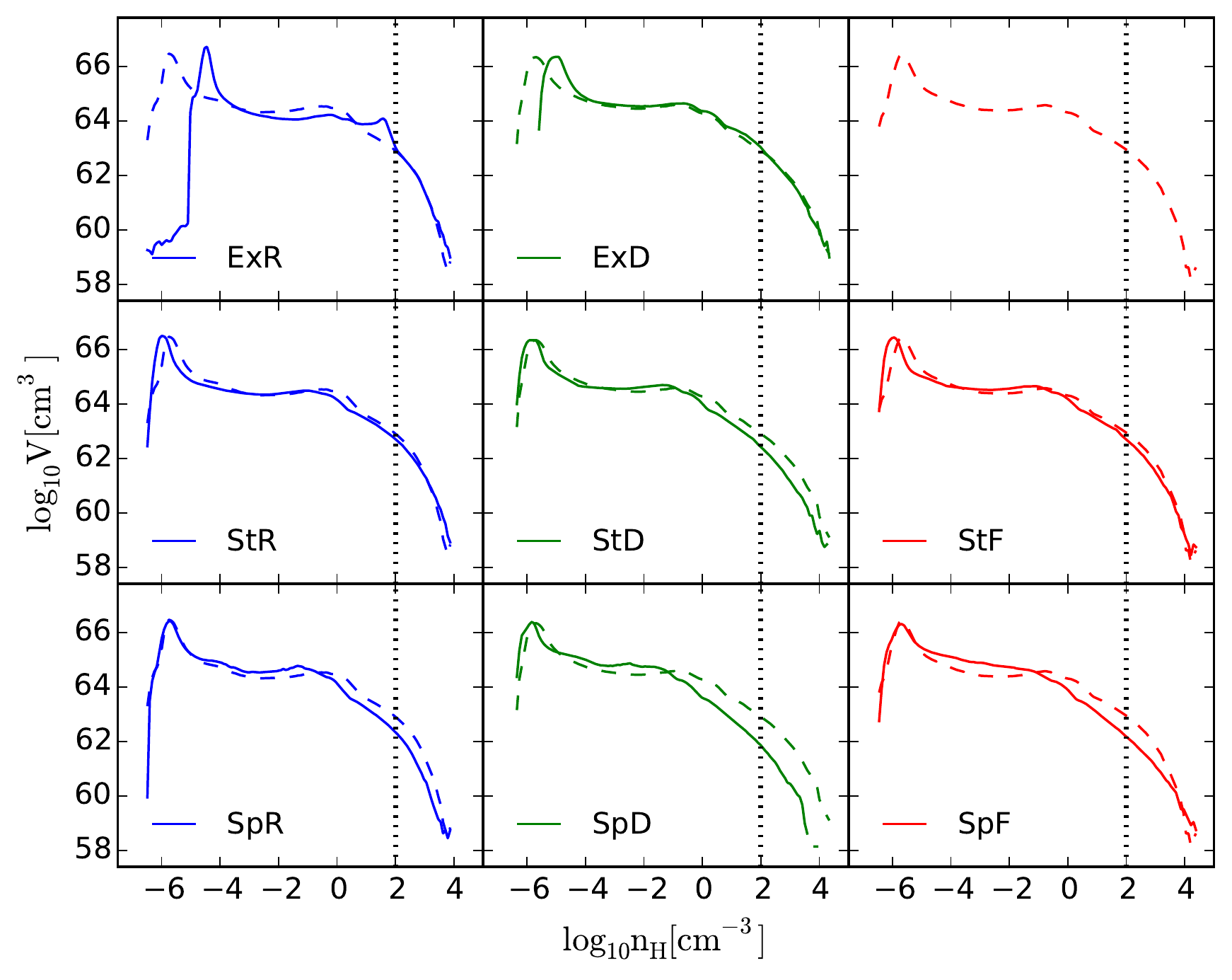}
	\caption{Volume-weighted PDF for gas density evaluated in an annulus of 2.5 kpc $< R < 8.5$ kpc and $|z| < 1$ kpc. The dashed lines show the fiducial models (FiR, FiD, FiF) for direct comparison. The vertical dotted line indicates the threshold of hydrogen number density for clouds $n_H=100$ $\textrm{cm}^{-3}$.}
	\label{pdf_vol}
\end{figure}

Another way to explore the ISM is to look at the one-dimensional probability distribution functions (PDF) for the gas. Figure \ref{pdf_vol} shows the volume-weighted PDFs for gas density, evaluated over a volume extending radially from 2.5 to 8.5 kpc and $\pm 1$ kpc above and below the disc mid-plane. The vertical line indicates the threshold of hydrogen number density for cloud identification ($n_H = 100 \textrm{cm}^{-3}$). In each panel we also show the {\it Fiducial} runs (FiR, FiD, FiF) as dashed lines. 

The {\it Fiducial} runs show most of gas volume is made of very diffuse gas (i.e. $n_H < 10^{-5} \textrm{cm}^{-3}$) with a log-normal tail of cold gas that fuels cloud formation, pushing down to densities of the order of $n_H = 10^{4} \textrm{cm}^{-3}$. As seen in Figure~\ref{phase_plots}, there is little total mass in this low density gas and it exists primarily above and below the disc. In keeping with both Figure~\ref{phase_plots} and Figure~\ref{projection_300}, the FiR has less high density cloud gas than FiD and FiF and more cold gas at lower densities, reflecting its more uniform fragmentation of smaller clumps, while FiD has the greatest star-forming material.

In the {\it Extreme} runs (top row of Figure~\ref{pdf_vol}) there is a shift in the very low density regions ($n_H < 10^{-5} \textrm{cm}^{-3}$) to higher peak densities, which is somewhat stronger in ExR than in ExD, consistent with the reduction of the small amount of HISM gas in phase plot. A tail of extremely low density gas occupying small volume in ExR, corresponds to a small amount of gas at a range of temperatures ($10^3{\rm K}\leq T\leq 10^7{\rm K}$) in the phase plot. A peak at near $n_H = 10^2 {\rm cm}^{-3}$ in ExR indicates the high density region near the centre of the disc, which is resisting fragmentation. ExD shows no discernible difference in the total volume of cloud gas, suggesting that while the outer disc may be better supported in the {\it Extreme} case, larger clumps are being formed in the inner disc. 

In the second row we compare the {\it Stellar Profile} gas density discs with the {\it Fiducial} runs. As with the previous phase plots, we see the extra gas and different profile does not affect the distribution within the ISM. There is only minimal difference between the two, with a negligible shift in the very low density ($n_H = 10^{-6} {\rm cm}^{-3}$) region, also seen in the HISM regions in phase plots. All {\it Stellar Profile} calculations show a slight decrease in the dense gas tail ($n_H > 1{\rm cm}^{-3}$), with StD being the most noticeable. The latter is from the increased stability in the outer disc suppressing the formation of clouds. As we will see more clearly in Section~\ref{Cloud_properties}, clouds in the {\it Stellar Profile} runs are larger, leading to looser outer envelopes that can be pulled apart by tidal interactions, reducing the volume of very dense gas. 

In the {\it Spiral} cases these is a slight rise in the volume of low density gas ($10^{-5} \textrm{cm}^{-3} < n_H < 1 \textrm{cm}^{-3}$), consistent with the expansion of the WISM seen in the phase plots. There is also a noticeable decrease in volume of dense gas, particularly the SpD calculation. This is due to the large interarm regions, as while the arms host comparatively fewer but larger and more massive clouds, they are not extended enough to counter the lack of clouds between arms.

%%%%%%%%%% subsection %%%%%%%%%%
\subsection{Cloud properties}
\label{Cloud_properties}

Moving away from exploring the continuous ISM, we next look at the properties of the individual clouds as defined in Section~\ref{Numerical methods}. These are the environments that will determine the star formation in a galaxy.

\subsubsection{Cloud mass and radius}
\label{cloudMR}

Figure~\ref{cloud_mass} shows the distribution of cloud masses detected within an annulus of $2.5{\rm kpc}\leq R \leq 8.5{\rm kpc}$. Each panel shows a comparison of the Ex, St and Sp models against their {\it Fiducial} counterparts (shown by dashed lines). 

For the {\it Fiducial} calculations, the typical cloud mass is approximately $10^{5.5}-10^6 {\rm M_\odot}$, with the maximum mass reaching just past $10^7$\,M$_\odot$. Allowing for an atomic envelope of similar mass to the molecular core, these values agree with observed GMC properties \citep{Rosolowsky2003, Roman2010, Fukui2009}. The lack of stellar feedback in these calculations also negates a significant mass loss mechanism from these collapsing clouds, which is why clouds masses are on the slightly larger side compared to observational studies \citep{Rosolowsky2007b}.
The {\it Rise} FiR simulation has the most uniform cloud mass, reflecting the even fragmentation into small clumps seen in Figure~\ref{projection_300}. The broadest profile is seen in FiD, due to the formation of larger clumps near the disc centre, compared with the sparse outer regions. 

The uniformity in typical cloud properties is notable, since the runs have quite different values for $\Gamma$. This is particularly of note in the FiR and FiF cases, which have similarly shaped mass profiles, only shifted to slightly lower masses in the FiR case, but moderately different values for $\Gamma$. However, the shear parameter, $S$, is similar in both runs, adding to the interpretation that this is the more relevant diagnostic of cloud properties. 

In the {\it Extreme} models (first row of Figure\;\ref{cloud_mass}), there is an increase in cloud mass in both the ExR and ExD cases compared to the {\it Fiducial} run, although the shift is far more significant in the ExR simulation. The peak of cloud mass distribution in ExR is now more ambiguous due to flattening the range of $10^5-10^7$\;$\textrm{M}_\odot$, with a lower population of low mass clouds, and a much greater high mass tail. This reflects the similar quantities of cloud gas seen in Figure~\ref{pdf_vol}, but the far smaller number of clouds listed in Table~\ref{cloudNo} (FiR: 1897 clouds, ExR: 945); the dense gas persists but is now forming more massive structures. The images in Figure~\ref{projection_300} reveal what has happened; with fragmentation restricted to the $Q$ unstable outer galaxy edge, clouds are closely packed which leads to mergers into bigger objects. 

The ExD run shows a slight reduction in the amount of clouds at typical cloud mass, with a slight widening of the distribution (to both high and low mass tails). This also reflects the {\it Extreme} environments ability to emphasise trends, with the fragmentation ratio of the inner and outer discs being more strongly marked compared to the {\it Fiducial} case.      
\begin{figure}
\centering
	\subfigure{
	\includegraphics[width=8.5cm]{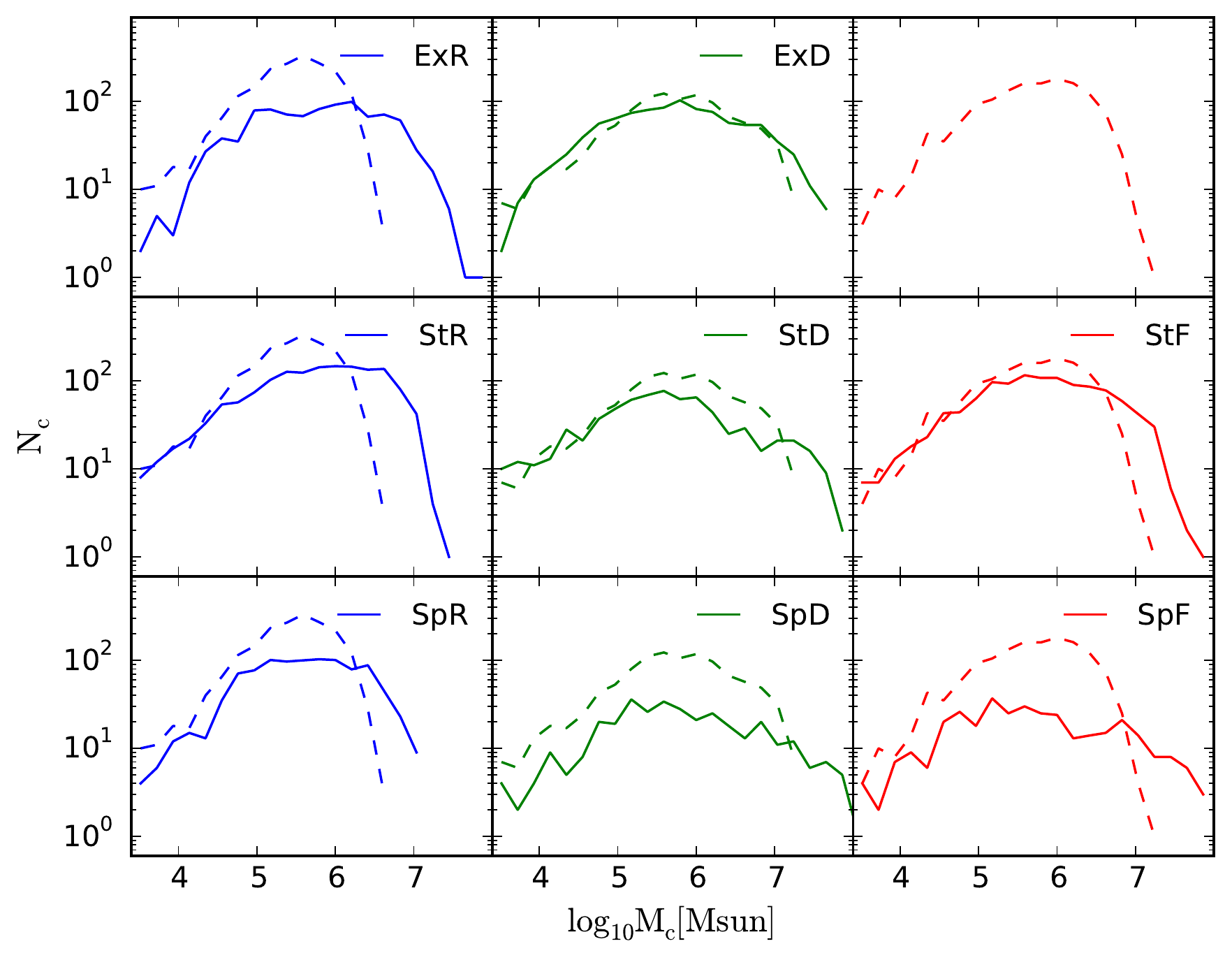}}
	\caption{Cloud mass distribution for the Ex, St and Sp models (solid lines) plotted against the of the {\it Fiducial} models (FiF, FiD, FiR: dashed lines). The typical cloud mass is roughly $10^{5.5}$ $M_\odot$--$10^{6.0}$ $M_\odot$. }
	\label{cloud_mass}
\end{figure}

The higher mass in the {\it Stellar Profile} models leads to a population of more massive clouds in all three runs. The low-mass end sees similar numbers of smaller clouds, suggesting that clouds are still born with a similar distribution of masses but the higher numbers result in the creation of more massive clouds through successive interactions. StR shows a decline in the quantity of typical clouds and a shift to the high mass tail, though it maintains the abrupt drop at a high mass limit, rather than a steady decline like the StD and StF curves. This is from the relatively uniform fragmentation of the {\it Rise} discs, which produce a more consistent collection of clouds, compared with the spread in sizes in StD and StF. 
    
Models with a spiral perturbation show the biggest difference across all three runs compared to the {\it Fiducial} calculations. They uniformly decrease the overall number of clouds at low and intermediate masses, and push up the peak cloud mass, flattening the mass distribution. These peak masses are similar to those of GMAs observed in the spiral arms in M51 \citep{Koda2009}, up to $10^8$ $\textrm{M}_\odot$, which also appears to be the upper limit across all models. These are likely created through successive mergers of clouds with a mass similar to the peak value in the {\it Fiducial} case, as seen in \citet{Fujimoto2014a}. The disc with the strongest spiral response, SpF (see Figure\;\ref{spiral_circ}), has the most flattened cloud mass profile, whereas the weakest spiral responses (SpR) still has an asymmetric mass distribution similar to ExR, StR and FiR. Conversely, the stronger response of SpD and SpF have made an almost symmetrical mass distribution. More massive star forming clouds being inherent to spiral arms is seen observations and simulations \citep{Koda2009,Dobbs2011b}, with more flocculent discs hosting lower mass clouds \citep{Rosolowsky2003}. The lack of any interarm structure in the strongest spiral (SpF) is likely the cause of the dearth of medium-low mass clouds, with a ``go big or go home'' response as upstream gas falls into the spiral shocks.
    
The use of rotation curves with high velocity maxima, gas tracing the stellar surface density, and spiral perturbations all encourage the production of massive clouds regardless of the shape of the rotation curve. The {\it Decrease} curves in particular show large changes compared to the {\it Fiducial} runs, even in the case of ExR where a large fraction of the disc is stable to fragmentation.

\begin{figure}
\centering
	\includegraphics[width=8.5cm]{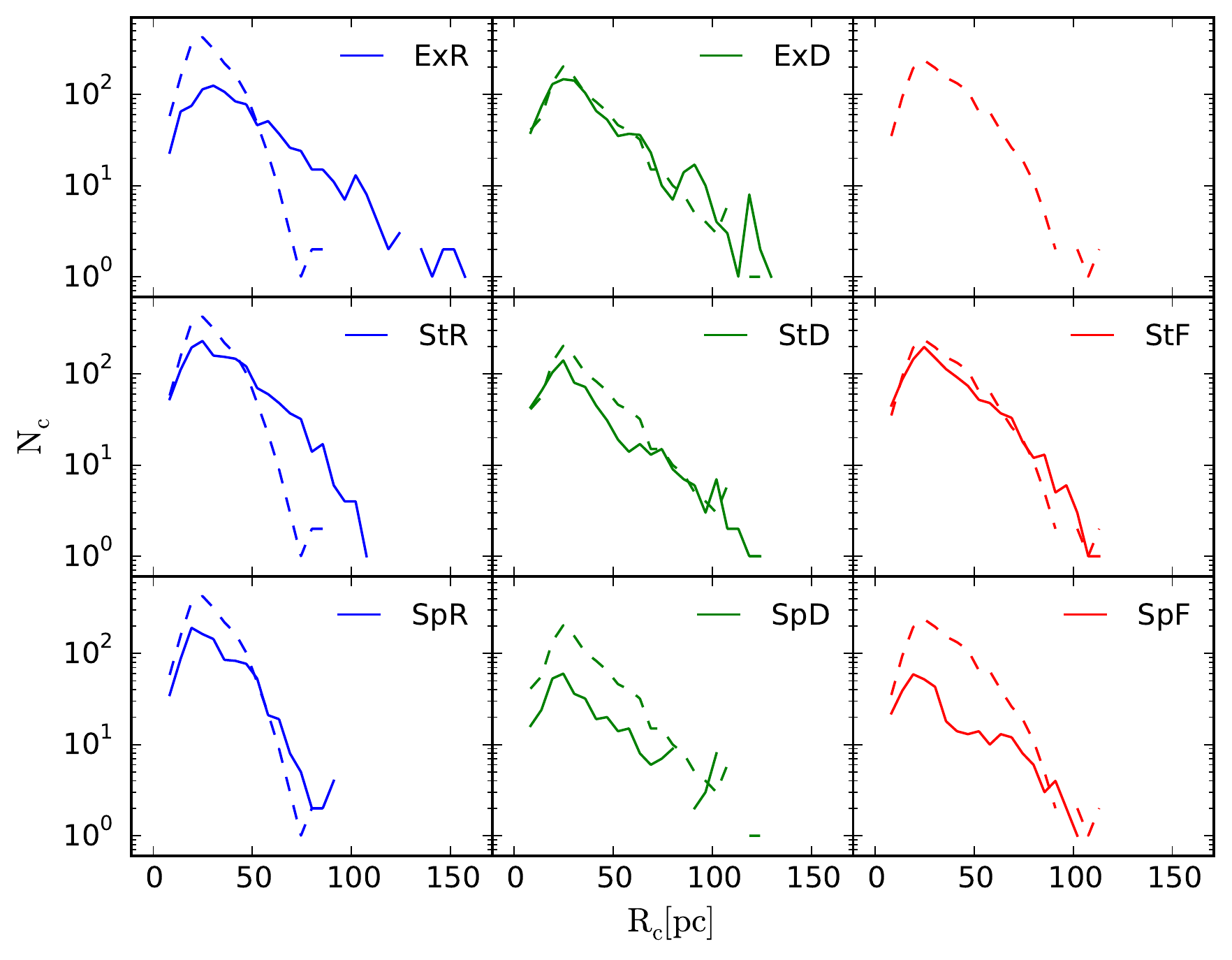}
	\caption{Cloud radius distribution for the Ex, St and Sp models (solid lines) plotted against the of the {\it Fiducial} models (FiF, FiD, FiR: dashed lines). The typical cloud radius is around 20pc.}
	\label{cloud radius}
\end{figure}

%%%-radius:
Figure \ref{cloud radius} shows the distribution of cloud radii across all models, again with the {\it Fiducial} cases plotted as dashed lines in each panel. The radius of each cloud is calculated as $R_c = \sqrt{(A_{\rm xy}^2 + A_{\rm yz}^2 + A_{\rm xz}^2)/3\pi}$, where $A_{\rm ij}$ is the projected area of the cloud in the $i-j$ plane for each dimension. In all cases the peak of the distributions tend to lie at approximately $R_c=20$pc, similar to observed clouds in the Milky Way \citep{Heyer2009}. Additionally, in Figure~\ref{larson_fig} we show the mass-radius relation for the clouds formed in these simulations, with the theoretical straight line fit from \citet{Roman2010} shown as dashed lines. 

As with other properties, such as density PDFs and cloud mass, the {\it Flat} rotation curves lies between the {\it Decrease} and {\it Rise}, with the latter having the most abrupt steep drop in cloud radii (with almost no clouds with $R_c>75$pc) due to the uniform disc fragmentation and the {\it Decrease} model having the broadest distribution from the ratio in fragmentation between inner and outer disc. This is clearly seen in Figure\;\ref{larson_fig}, where the FiR clouds take up a much smaller region of the $M_c-R_c$ parameter space compared to FiF and even more so compared to FiD.

The {\it Extreme} ExR curve shows a huge increase in the radii of clouds compared to FiR, pulling away from the steep drop to radii of up to 150pc. This corresponds with the much higher mass seen in Figure~\ref{cloud_mass}. ExD on the other hand looks similar to FiD, with only a small increase in the number of large clouds. Figure\;\ref{larson_fig} shows the clouds have a wider distribution of surface densities, but still follow the same trend as in the {\it Fiducial} case. ExD forms some of the largest radii clouds in the {\it Decrease} curves, but the cloud mass distribution does not reach as high peaks as SpD or SpF calculations. This can also be seen in Figure\;\ref{larson_fig}, where the largest clouds in the ExD case sit lower (with lower surface density) than the largest clouds in StD and SpD. 

The StR model also shows a larger difference between the {\it Fiducial} run than StD and StF. Interestingly the StD model clouds seem to have gained relatively more mass than volume, as they have pulled up and away from the straight line fit, following a steeper gradient than the {\it Rise} models. The added shear in the Keplerian-like galaxy rotation could increase tidal interactions between clouds, stripping low density outer layers to produce more compact clouds. The more solid-body like rotation curve on the other hand, has lower shear values and the clouds can ``puff-up'' to retain larger radii at equivalent masses without having their peripheries sheared away. 

Similarly to mass, the {\it Spiral} perturbation smooths out the radius distribution to larger radii with a reduced population of low-medium mass clouds. This results in a population of, on average, larger and more massive clouds but fewer in number. As with the StR run, SpR has only a small increase to large radii compared to mass, resulting in a $M_c$--$R_c$ population that sits on the dashed line fit in Figure\;\ref{larson_fig}. Meanwhile, the SpD and SpF have an extended tail of high mass, high radii clouds that are slightly more compact that the main body of clouds present in the {\it Fiducial} runs. The bunching-up of clouds in the spiral arm could have a similar effect as shear to allow the outer envelop of the clouds to be lost via tidal stripping. 

\begin{figure}
\centering
\resizebox{1.0\hsize}{!}{\includegraphics[trim = 10mm 10mm 0mm 0mm]{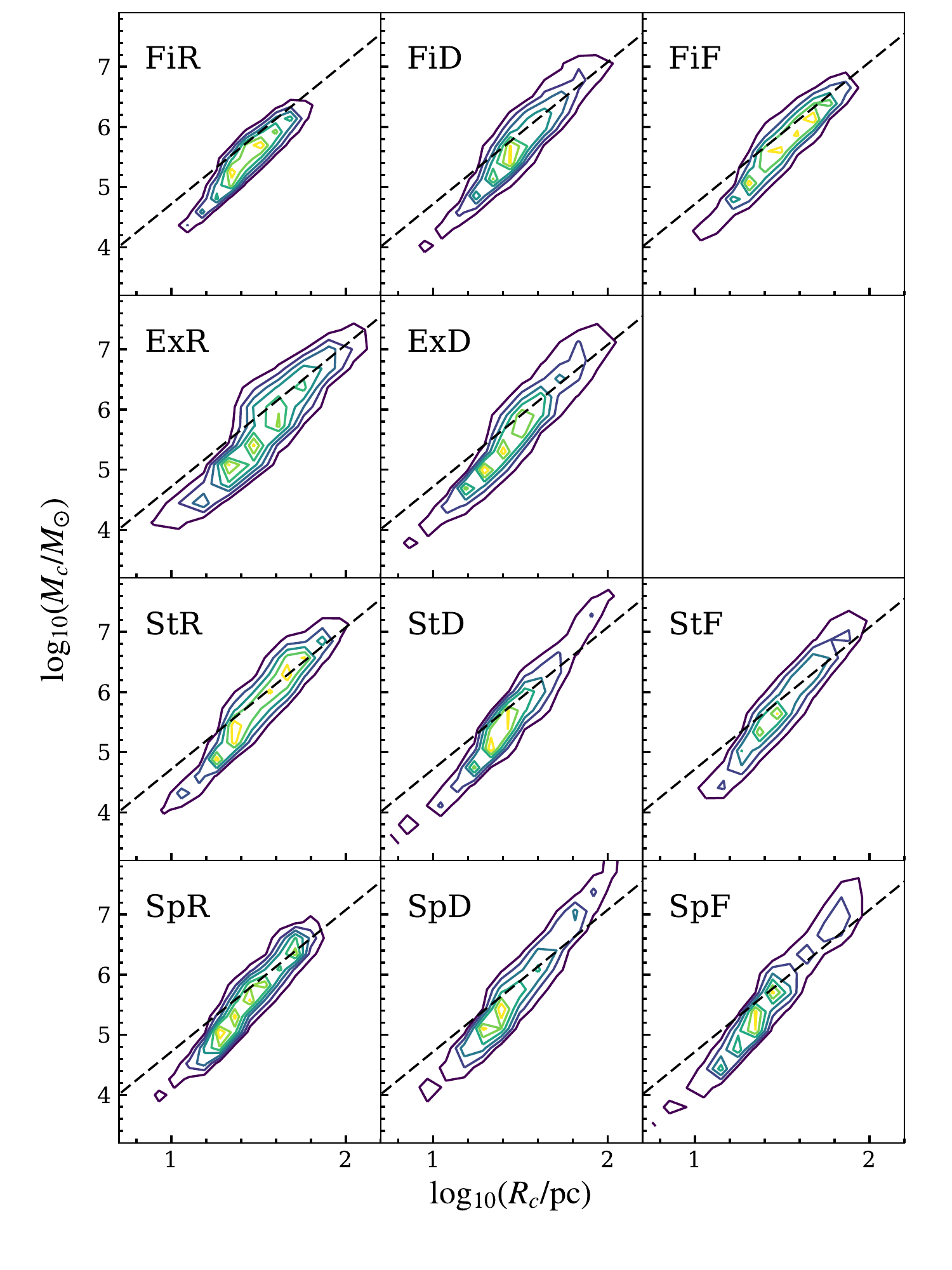}}
	\caption{The mass-radius relation for the clouds in each simulation presented in this study. Cloud properties have been binned into logarithmic $M_c$--$R_c$ bins, with the theoretical straight line fit from \citet{Roman2010} shown as the dashed line in each panel.}
	\label{larson_fig}
\end{figure}

%%%-vel dispersion-%%%
\subsubsection{Cloud stability}

In order to quantify the stability of clouds in our simulations, we plot the 1D velocity dispersion in the clouds ($\sigma_c$) and the virial parameter ($\alpha_{\rm vir}$) in Figures \ref{cloud_sigma} and \ref{cloud_alpha} respectively. $\alpha_{vir}$ is defined as:
\begin{equation}
\alpha_{\rm vir} = \frac{5\sigma_c^2R_{c}}{GM_c}
\end{equation}
where $\sigma_c$ is the mass-averaged one-dimensional velocity dispersion of the cloud, i.e., $\sigma_c \equiv (c_s^2+\sigma_{nt,c}^2)^{1/2}$, where $\sigma_{nt,c}$ is the 1D rms velocity dispersion about the cloud's centre-of-mass velocity. The virial parameter of a molecular cloud describes the ratio of internal, supporting energy to its gravitational energy, with a value of 1.0 indicating a virialised system.

The typical value of $\sigma_c$ across all models is approximately 2--3 $\textrm{km\;s}^{-1}$. High 1D velocity dispersion tails are evident in all cases compared to the {\it Fiducial} runs, though this only appears to drive a slight increase in the virial parameter, which appears to be compensated by the steeper increase in cloud masses in these models (see Figure \ref{cloud_mass}). The velocity dispersion shows the same trend with cloud mass distribution, with {\it Extreme} models showing the largest difference in the high 1D velocity dispersion to the {\it Fiducial} runs. The {\it Stellar Profile} calculations mainly drive an increase in the high dispersion tail, while the {\it Spiral} runs seem especially effective at reducing the low/medium dispersion clouds up to higher values, rather than simply adding to the high dispersion tail. The truncation in velocity dispersion for the {\it Rise} SpR calculations appears to be earlier for $\sigma_c$ than for $M_c$ or $R_c$, which appear more similar to their {\it Decrease} SpD and {\it Flat} SpF counterparts. No clouds are seen with $\sigma_c>18$ $\textrm{km}^{-1}$ in any of the {\it Rise} models, whereas {\it Flat} and {\it Decrease} models have values peaking at almost double this value ($\approx 32 {\rm km\,s^{-1}}$ in SpD).

\begin{figure}
\centering
	\subfigure{
	\includegraphics[width=8.5cm]{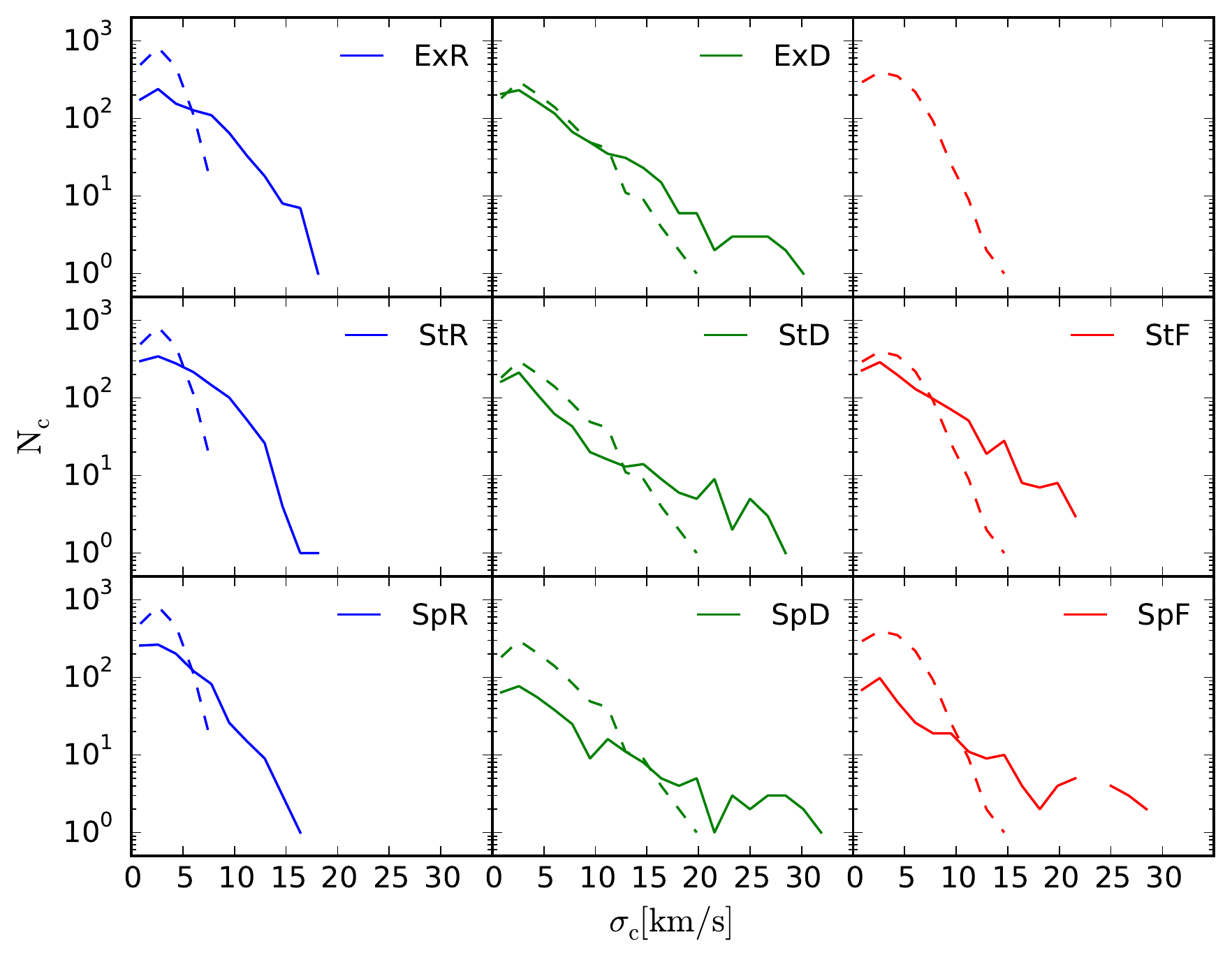}}
	\caption{Distribution of 1D velocity dispersion for clouds ($\sigma_{c}$) formed in the Ex, St and Sp models (solid lines) plotted against the of the {\it Fiducial} models (FiF, FiD, FiR: dashed lines). The typical value is approximately $2-3{\;\rm km\;s}^{-1}$ across all models.}
	\label{cloud_sigma}
\end{figure}

\begin{figure}
\centering
	\subfigure{
	\includegraphics[width=8.5cm]{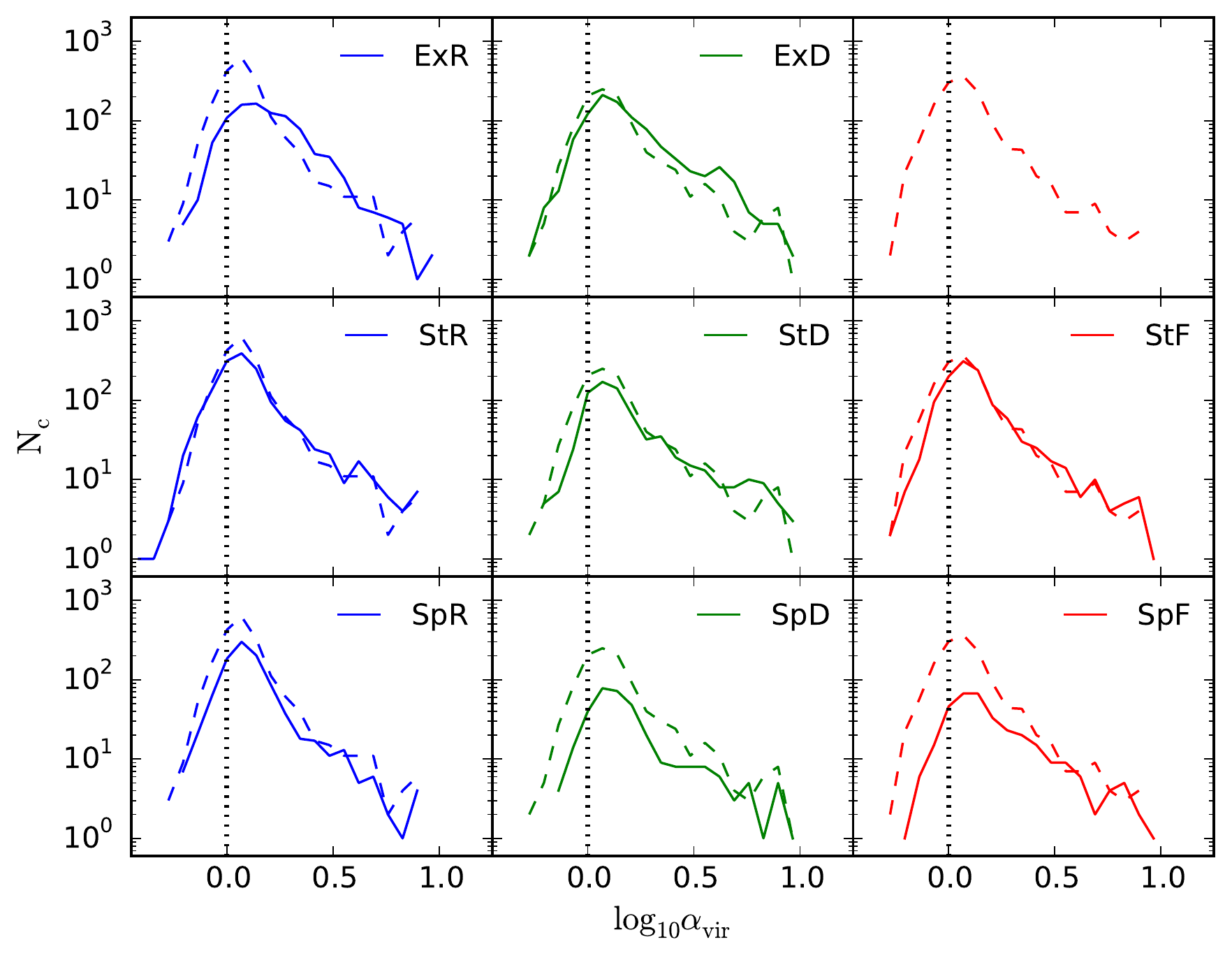}}
	\caption{Distribution of virial parameter, $\alpha_{vir}$ for the Ex, St and Sp models (solid lines) plotted against the of the {\it Fiducial} models (FiF, FiD, FiR: dashed lines). The vertical dotted line indicates the $\alpha_{\rm vir} = 1$ limit.}
	\label{cloud_alpha}
\end{figure}

The distribution of virial parameter in all case peaks at around $\alpha_{\rm vir}=1$, implying clouds are borderline bound. There is a substantial population of clouds with $\alpha_{\rm vir}>1$ (and even with values $>2$) indicating a significant unbound population. It has been documented in numerous studies that clouds can exist with $\alpha_{\rm vir}>1$ both in simulations \citep{TaskerTan2009,Ward2016} and observations \citep{Rosolowsky2007a,Bolatto2008,Heyer2009}. Simulations of \cite{Dobbs2011} suggest that while in general clouds may be unbound, it is the denser cores and inner regions that are becoming dense enough to form stars, thus the whole cloud structure need not necessarily be bound, only the inner regions. There is also increasing evidence that $\alpha_{\rm vir}$ is a poor descriptor of the dynamical state of a GMC \citep{Baba2017} and it should be noted that it is a particularly difficult parameter to measure observationally \citep{Pan2015, Pan2016}.

Values of $\alpha_{\rm vir}$ appear relatively insensitive to rotation curves and gas distribution used, with most distributions in Figure 
\,\ref{cloud_alpha} appearing identical. The fact that St and Fi runs are identical highlights that the exact gas distribution and gas reservoir have no influence on cloud stability. ExR is the main exception, which has the largest cloud sizes but similar values of $M_c$ and $\sigma_c$ to other models, resulting in a peak of the $\alpha_{\rm vir}$ distribution in the unbound regime. As stated in Sec.\;\ref{cloudMR}, this may be due to the very low shear in ExR, allowing clouds to maintain a loosely bound periphery compared to their higher shear brethren.

The {\it Spiral} runs show the greatest departure from their {\it Fiducial} counterparts, with SpD and SpF reducing the population of bound/overbound clouds and increasing their unbound population. This is likely driven by a combination of $M_c$ and $\sigma_c$ (see also the clouds in SpD and SpF strongly pulling up and away from the dashed line in Figure\;\ref{larson_fig}). \citet{Dobbs2011b} also found that calculations with spirals tend to form less bound clouds than the axisymmetric cases, possibly because a spiral can sweep gas into clouds independently of gravity, making $\alpha_{\rm vir}$ a poorer indicator of cloud stability.

%%%%%%%%%% Conclusion %%%%%%%%%%
\section{Conclusions}
\label{Conclusions}

We modelled a set of eleven isolated galaxy discs with different global environments set by their background potential. The overall fragmentation structure of the discs were sensitive to the potential environment, dominated by both the Toomre $Q$ parameter for gravitational stability and the disc shear parameter, $S$. Our {\it Extreme} rotation environments, which see circular velocities up to 500\,km\,s$^{-1}$, demonstrate the places where each regime reigns; in the ExR (Extreme Rising) simulation, the high epicycle frequency of the inner region drives up $Q$ to prevent collapse. However, in the outer parts of the ExD (Extreme Decreasing) disc, shear acts against collapse to produce a sparser collection of clouds. The background potential also determined the response of the disc to an applied spiral perturbation. Discs with rising velocity curves correspond to dark matter dominated discs whose mass overrules the spiral model excitation to result in a weaker generated spiral in the gas.

Despite these differences in disc structure, the properties of a typical cloud were reasonably robust, sitting around $10^{5.5}-10^{6.0}$\,M$_\odot$ and 20\,pc for most runs and following the observed trend of the size-mass relation. However, the distribution tails containing the largest and smallest clouds in the population were sensitive to the environment, varying both with the change in rotation curve and the application of the spiral potential. Environments that offered greater stability or more shear support, such as those in the {\it Extreme} rotation curves, produced a tail of larger clouds. Similarly, increasing the cloud number density either through gathering clouds into a spiral shock or using a heavier disc, also led to a boost in cloud size due to an increased merger rate. Lower levels of shear across the galactic disc such as those in the {\it Rise} curves produced a uniform fragmentation that led cloud profiles concentrated around the peak values, with smaller tails.   

Our models have explored the external effect of galactic potential on the evolution of the star-forming clouds. The evolution of these clouds will also be influenced by the internal processes of star formation and feedback. Exactly how stellar production affects GMCs is not yet well understood (see e.g. \citet{Dobbs2014b,Padoan2014}). Previous studies using an identical set-up to our fiducial flat (FiF) disc have explored this in a series of simulations that gradually increased the included stellar physics \citep{TaskerTan2009, Tasker2011, Tasker2015}. These works indicated that high mass tail is reduced as gas is converted into stars, but typical cloud properties remain unchanged by their internal stellar processes. This suggests that the main driver for GMC evolution is external forces such as gravity and shear, rather than star formation and feedback. Since star formation is governed by the GMC properties and our typical GMC is constant among our discs, we do not expect our result to be strongly influenced by the introduction of additional stellar physics.

Another section of the parameter space concerns the generation of the galactic spiral. Since we wished to explore how shear impacts cloud growth, we elected to include steady-state density wave in our simulations that would be uniformly maintained over a long period, represented as a rigidly rotating potential. However, different spiral structures would be induced by the inclusion of a live stellar disc that result in transient and dynamic spiral modes have been shown contradictory to to classical density wave spirals \citep{Dobbs2010, Grand2015, Baba2016}. Exploration of how GMC properties are influenced by different non-axisymmetric features (e.g. dynamic arms, tidal spirals, inner bars) in galactic discs is an interesting problem and the focus of future work. Spiral arms formed in live stellar discs may produce clearer differences than seen with an analytic background potential. For instance, the Fall model would be expected to show fewer and stronger arms than Rise, which may bring with it an amalgamation of larger and shorter-lived GMC complexes as they build up within and shear apart upon leaving the stronger spiral arms.

Overall, the strongest impact on the cloud population was made by the inclusion of the spiral potential. This swept cloud gas into the spiral arms, producing GMA-sized structures with broader profile of properties than for the flocculent runs.

\section*{Acknowledgments}
We used the \textsc{yt} package \citep{Turk2011} for parts of our analysis and plotting of data. Numerical computations were carried out on Cray XC30 at the Center for Computational Astrophysics (CfCA) of the National Astronomical Observatory of Japan. Takashi Okamoto acknowledges the financial support of The Ministry of Education, Culture, Sports, Science and Technology (MEXT) KAKENHI Grant (16H01085).

\bsp
\label{lastpage}
\end{document}